\documentclass{jfm_arxiv}

\usepackage{graphicx}
\usepackage{newtxtext}
\usepackage{newtxmath}
\usepackage{natbib}
\usepackage{hyperref}
\hypersetup{
    colorlinks = true,
    urlcolor   = blue,
    citecolor  = black,
}

\newcommand{\RomanNumeralCaps}[1]
\linenumbers

\usepackage{amsmath}
\usepackage{amssymb}

\usepackage{xcolor}
\definecolor{nintendo_blue}{RGB}{62,126,192}
\definecolor{nintendo_yellow}{RGB}{236,194,11}
\definecolor{nintendo_red}{RGB}{211,70,32}
\definecolor{nintendo_green}{RGB}{69,191,148}

\definecolor{matlab_ao}{rgb}{0,0.4470,0.710}
\definecolor{matlab_daidai}{rgb}{0.8500,0.3250,0.0980}
\definecolor{matlab_ki}{rgb}{0.9290,0.6940,0.1250}
\definecolor{matlab_murasaki}{rgb}{0.4940,0.1820,0.5560}
\definecolor{matlab_macha}{rgb}{0.4660,0.6740,0.1880}
\definecolor{matlab_sora}{rgb}{0.3010,0.7450,0.9330}
\definecolor{matlab_azuki}{rgb}{0.6350,0.0780,0.1840}

\usepackage{MnSymbol}

\usepackage{subcaption}

\usepackage{tikz}
\usetikzlibrary{calc}

\usepackage{array}
\usepackage{multirow}

\newcommand{\superficialavg}[1]{\left\langle#1\right\rangle_{\mathrm{s}}}
\newcommand{\intrinsicavg}[1]{\left\langle#1\right\rangle_{\mathrm{i}}}

\newcommand{\porosity}{\epsilon}

\newcommand{\permeability}{K}

\newcommand{\tauinv}{\tau_{\mathrm{inv}}}

\newcommand{\tauvisc}{\tau_{\mathrm{visc}}}

\newcommand{\tauen}{\tau_{\mathrm{en}}}

\newcommand{\Hg}{\mbox{\textit{Hg}}}

\renewcommand{\Vec}[1]{\boldsymbol{#1}}

\newcommand{\steady}{\mathrm{steady}}

\title{Similarity of start-up flow in porous media for large pressure gradients}

\author{Yoshiyuki Sakai\aff{1}
  \corresp{\email{yoshiyuki.sakai@tum.de}},
  Lukas Unglehrt\aff{1,2}
 \and Michael Manhart\aff{1}}

\affiliation{
\aff{1}
Technical University of Munich, TUM School of Engineering and Design, TUM Department of Civil, Geo and Environmental Engineering, Arcisstr. 21, 80333 Munich, Germany
\aff{2}
UNGLEHRT GmbH \& Co. KG Bauunternehmen, Allgäuer Str. 31, 87700 Memmingen, Germany
}
\begin{document}
\maketitle

\begin{abstract}
We investigate the start-up flows through ordered porous media (hexagonal close-packed, face-centred cubic and body-centred sphere packs) by means of direct numerical simulations. 
The flows are initiated from rest and driven by a constant pressure gradient, allowing us to examine the transient development across a wide range of Hagen numbers. 
Dimensional analysis identifies two relevant time scales: the viscous diffusion time $\tauvisc$ and the inviscid time $\tauinv$. 
While the small-time behaviour follows the viscous asymptotics of Johnson et al. [J. Fluid. Mech. 176, 379 (1987)], the subsequent emergence of nonlinear effects is universally governed by the inviscid time $\tauinv$, rather than by any critical Reynolds number.

At the pore scale, the transient evolution is characterised by the growth of thin vorticity layers on the sphere surfaces, their detachment into the pore space around $t \sim \tauinv$, and the formation of inertial cores. 
Despite geometric differences, these processes occur in a remarkably similar sequence across all three packings. 
Vorticity magnitude exhibits laminar boundary-layer scaling with Hagen number, while in the body-centred cubic sphere pack case a transition towards turbulent-type scaling is observed.

These results establish $\tauinv$ as a unifying measure for the onset of nonlinearity in strongly accelerated porous media flows, with direct implications for the modelling of unsteady transport in natural and engineered systems.
\end{abstract}

\section{Introduction} 
\label{sec:intro}

Unsteady flow through porous media plays a fundamental role in numerous natural and engineered systems.
In nature, nutrient and gas transport within coral reef skeletons are governed by unsteady currents driven by waves, tides, and storm events \citep{Lowe2008a,Rogers2016}.
Such hydrodynamic events, as well as density-driven circulation, also drive unsteady flows in coastal aquifers and intertidal zones, producing highly transient flow patterns in permeable sediments and subterranean estuaries \citep{Robinson2007,Santos2012}. 
Frequent storm surges can also cause persistent unsteady conditions in coastal groundwater systems, preventing full recovery between events and inducing repeated perturbations in subsurface water table dynamics \citep{Nordio2023}.
In subsurface engineering, unsteady porous media flows are central to operations such as Carbon Capture and Storage (CCS), where supercritical CO$_2$ is injected into deep saline aquifers or depleted hydrocarbon reservoirs. These processes impose a sudden pressure gradient on the porous medium, initiating a transient flow regime \citep{Class2009,Sacconi2020}. 
    
The above examples demonstrate the prevalence of unsteady porous media flows. 
However, many theoretical and empirical models of porous media flow are based on steady or quasi-steady assumptions. 
This oversimplification often results in a high degree of predictive uncertainty by failing to account for the transient, nonlinear and sometimes chaotic nature of the flow during critical operational phases.
Instead, those established theories of the stationary flows should be viewed as baselines to which the temporally developing flows will eventually evolve into.
Subsequently, some prototypical transient flow problems need to be investigated for identification of the asymptotic behaviours of unsteady flow that any physically consistent model should reproduce. 
For instance, oscillatory flows due to a sinusoidal pressure wave have been frequently studied for the flow response to the high and the low frequency limits \citep[e.g.][]{Zhu2016,Unglehrt2022,Unglehrt2023,Unglehrt2024}, whilst the small- and large-time asymptotics can be determined in the start-up flow under a constant forcing.
Once these asymptotics are determined, investigations of the intermediate behaviours should follow to complete the overall picture.

Generally, those fundamental investigations are performed based on idealised porous media geometries to eliminate the geometric uncertainties.
Consequently, monodisperse sphere pack setup is often preferred to analyse the complex pore-scale flow dynamics in a laboratory-like controlled environment.
Examples of such investigations by means of experiments are \cite{Dybbs1984,Horton2009}, whist the numerical counterparts are \cite{Maier1998,HILL2001a,HILL2001,HILL2002a,Nguyen2018,He2019}.
Here, we follow this established approach by numerically investigating start-up flows through monodisperse sphere packs to extract mechanistic insights.

%

The rest of this introduction is organised as follows: First, we summarise the main features of stationary flows through porous media, which will be followed by an overview on the start-up internal flow with a special focus on the small-time asymptotic behaviour, the state-of-the-art of the nonlinear transient flows within porous media, and finally the research questions and objectives will complete this introduction. 

A stationary flow through a porous medium exhibits different flow states, namely: linear (Stokes), steady nonlinear, unsteady nonlinear, and turbulence-like chaotic flow \citep{Dybbs1984}.
Those flow states are determined by a dimensionless flow parameter called \textit{Reynolds number} $\Rey$.
As mentioned, stationary flows through porous media generally represent the large-time asymptotic of the corresponding unsteady flows.
It is, therefore, expected that weakly transient flows behave similarly to the stationary flows such that they can be described simply according to the instantaneous Reynolds number, given that the dimensional number slowly relaxes towards the steady-state level. 
In such scenarios, the flow has enough time to adapt to the new equilibriums at each instance.
Conversely, under strong acceleration or deceleration, the transient flow is unlikely to depend on the instantaneous Reynolds number.

The linear stationary flow occurs in the very low Reynolds number limit $\Rey \ll O(1)$, and is characterised by Darcy's law: a linear dependence of the superficial (volume-averaged) velocity on the driving body force and the counteracting viscous resistance.
In this flow regime, if present in the porous media, the fore-aft symmetry in the pore-scale velocity distribution is maintained.
%
In case of the nonlinear flow with a significantly higher Reynolds number $\Rey > O(1)$, an additional drag force arises from the nonlinear (inertial) term in the Navier--Stokes equations, and the nonlinear drag becomes progressively more dominant with increasing $\Rey$. 
%
%
At this stage, the fore-aft symmetry in the pore-scale velocity distribution is broken in the form of flow separations \citep{Sakai2020}.
Despite the increased complexity of the flow features, the nonlinear flow is still steady if the flow Reynolds number is moderate, whilst the flow structures maintain the spatial symmetries in the cross-stream directions imposed by the pore geometries \citep{HILL2002a,Sakai2020}.
%
Once the flow Reynolds number exceeds a certain threshold at $\Rey \sim O(10^2)$, the nonlinear flow finally becomes unstable through Hopf bifurcation, and eventually transitions to the chaotic flow state with $\Rey > O(10^2)$.

Generally, the development of wall-bounded start-up flow is also well-understood.
When a resting fluid is suddenly set into motion by a constant pressure gradient, the resulting velocity field is almost exclusively controlled by the pressure force, and an inviscid and irrotational potential flow combined with a boundary-layer characteristics emerges instantly.
In the context of transient flows through sphere packs, the above implies co-existence of the potential core flow and thin Stokes boundary layer around the sphere surfaces \citep{HILL2001}.
This boundary layer is initially infinitesimally thin, whose thickness $\delta$ grows with time as $\delta \sim \sqrt{\nu t}$, where $t$ is the elapsed time from the start-up whereas $\nu$ is kinematic viscosity of the fluid \citep{Schlichting.2017}.
Based on this boundary layer theory, the exact thin boundary layer asymptotics for porous media flow was derived by \citet{Johnson.1987} in the frequency domain, which can be expressed in the temporal domain by means of inverse Fourier transform \citep[][\S B.2.]{Unglehrt2023}.
Consequently, the small-time asymptotics of the start-up flow through porous media can be determined.

The growing boundary layers inside porous media eventually fill the entire pores at $t \rightarrow \infty$, and the flow fields converge to the steady-state distributions which represent the large-time asymptotes.
In case of linear start-up flows, the superficial velocity relaxes in an exponential manner towards the steady-state level after the aforementioned small-time asymptotic phase. 
During this process the flow field behaves like in a quasi-steady-state, as it was shown by \cite{HILL2001} and \cite{Zhu2014} who simulated such flows through simple cubic and random arrays and hexagonal closed pack (HCP) spheres, respectively.
%
Furthermore, this late-time exponential relaxation is known to have a characteristic time constant called \textit{energy time scale} $\tauen$, which makes the relaxations across different porous media similar.


The development of nonlinear transient flows through porous media is much less trivial, and it is known to differ qualitatively from the linear counterpart.
For instance, in contrast to the aforementioned exponential relaxation, an overshoot of the superficial velocity briefly before reaching to the steady-state was observed in the nonlinear flows through face-centred cubic (FCC) and the HCP sphere packs \citep{HILL2002a,Zhu2016,Sakai2020}.
During the flow development under strong enough acceleration, a sequence of flow structures develops inside the pores, and the structural evolution was found to be identical across steady/unsteady/chaotic nonlinear flow regimes.
Furthermore, it was shown that both the velocity overshoot and the pore-scale flow evolution manifest independent of their instantaneous and steady-state Reynolds numbers, and their formation times cannot be described by the normalisation based on the fluid viscosity and a fixed length, such as the sphere diameter \citep{Sakai2020}.
This implies that the quasi-steady-state assumption does not hold in the strongly nonlinear flow since the flow Reynolds number does not play a decisive role.
This observation is consistent with the finding of \citet{Zhu2016a} that the deviation from linearity is not a function of the instantaneous Reynolds number in the strongly accelerating porous media flow, which led to an alternative Reynolds number definition based on the Stokes boundary layer thickness.
This newly proposed instantaneous Reynolds number showed a notable improvement over more conventional fixed-length-based definition to predict the onset of nonlinearity in a range of porous media, however not conclusive. 
Finally, our more recent contribution \citep{Sakai2022} showed that the formation time of the velocity overshoot in the strongly accelerated flow through the HCP sphere pack converges to a constant multiple of so-called inviscid time scale $\tauinv$.
The generality as well as the physical mechanism behind of this finding are, however, still to be investigated.

In summary, the start-up flows through porous media driven by small pressure gradients can be generally well-explained by a combination of the small- and the large-time asymptotes. 
Our understanding in the development of the strongly accelerated start-up flows, on the other hand, is much less established.
In particular, the onset process of nonlinearity in such flows, emerging in between the small- and the large-time limits, has not been explored extensively to-date.
Besides obvious dynamical interests, addressing this knowledge gap is also crucial for the modelling of the unsteady flow, however, parametrisation of the nonlinear effects is a non-trivial task \citep{Unglehrt2024}.


The above survey leads to the following questions for this study, namely: 
(i)
How general is $\tauinv$ as a temporal measure to describe strongly accelerated nonlinear porous media flow?
For instance, can it be used to describe other nonlinear flow through different sphere pack geometries?
(ii) The materialisation of the velocity overshoot which is timed by $\tauinv$ is a nonlinear phenomenon; Does this mean that $\tauinv$ generally times the onset of nonlinearity? If so, what is the physical mechanism behind?
(iii) Which of the changes in the flow field are associated with $\tauinv$?

Consequently, we define the following objectives and the corresponding structure of this paper.
We analyse the high-fidelity Direct Numerical Simulation (DNS) dataset of flow through the HCP spheres from \cite{Sakai2020}, as well as two additional DNS datasets of flow through FCC, and body-centred cubic (BCC) sphere packs.
After presenting the simulation approach and the properties of the datasets,
we perform a dimensional analysis to determine the governing time scales of the start-up flow through sphere packs.
Then, based on the boundary layer theory and an order of magnitude analysis, we show that the onset of the nonlinearity emerging from the sphere surfaces should be timed by $\tauinv$.
The validity of the above conjecture is then demonstrated based on the DNS datasets.
At the end, by means of scaling arguments, it becomes evident that the employed boundary layer theory in the start-up flow can consistently describe the considered transient flows through sphere packs, not only qualitatively but also quantitatively.  

\section{Numerical methods} 
\label{sec:numerics}

The flow solver MGLET was used to generate all three DNS datasets being considered in this study.
The code solves the following incompressible Navier-Stokes equations:

\begin{align}
    \bnabla \bcdot \boldsymbol{u} &= 0\\
    \partial_t \boldsymbol{u} + \bnabla \bcdot (\boldsymbol{u} \otimes \boldsymbol{u}) &= 
    - \frac{1}{\rho} \bnabla p + \nu \Delta \boldsymbol{u} -\frac{1}{\rho} \bnabla\! \intrinsicavg{p}\boldsymbol{e}_x\ ,
    \label{eqn:nse-dimensional}
\end{align}

\noindent
where the velocity vector $\boldsymbol{u} = [u, v, w]$ corresponds to the Cartesian spatial directions $\boldsymbol{x} = [x,y,z]$, 
whereas $p$ is the pressure deviation from the volume-averaged level (i.e. $\left\vert\bnabla\!\intrinsicavg{p}\right\vert\cdot x$,
with $\left\vert\bnabla\!\intrinsicavg{p}\right\vert$ being the intrinsic pressure gradient that drives the flow).
Moreover, $\rho$ and $\nu$ are the fluid density and kinematic viscosity respectively.

The above velocities are stored in staggered arrangement and discretised in space by the second-order central finite-volume method, which is combined with 
a second-order mass-conserving immersed boundary method \citep{Peller2006,Peller2010,Sakai2020} to handle arbitrary-shaped domains.
An explicit third-order low-storage Runge-Kutta method \citep{Williamson1980} integrates the governing equations in time, whilst the projection method of 
\cite{Chorin1968} decouples the velocity and the pressure computations.
For every Runge-Kutta sub-steps, a pressure Poisson equation is solved by means of a Strongly Implicit Procedure (SIP) solver \citep{Stone1968}, which has been recently
optimised for the modern SIMD processors.
The code is written in Fortran and parallelised by means of Message Passing Interface (MPI), which scales efficiently
up to $O(10^5)$ parallel processes on the modern massively-parallel systems \citep{Sakai2019}.
\section{Direct numerical simulation datasets}
\label{sec:dataset}
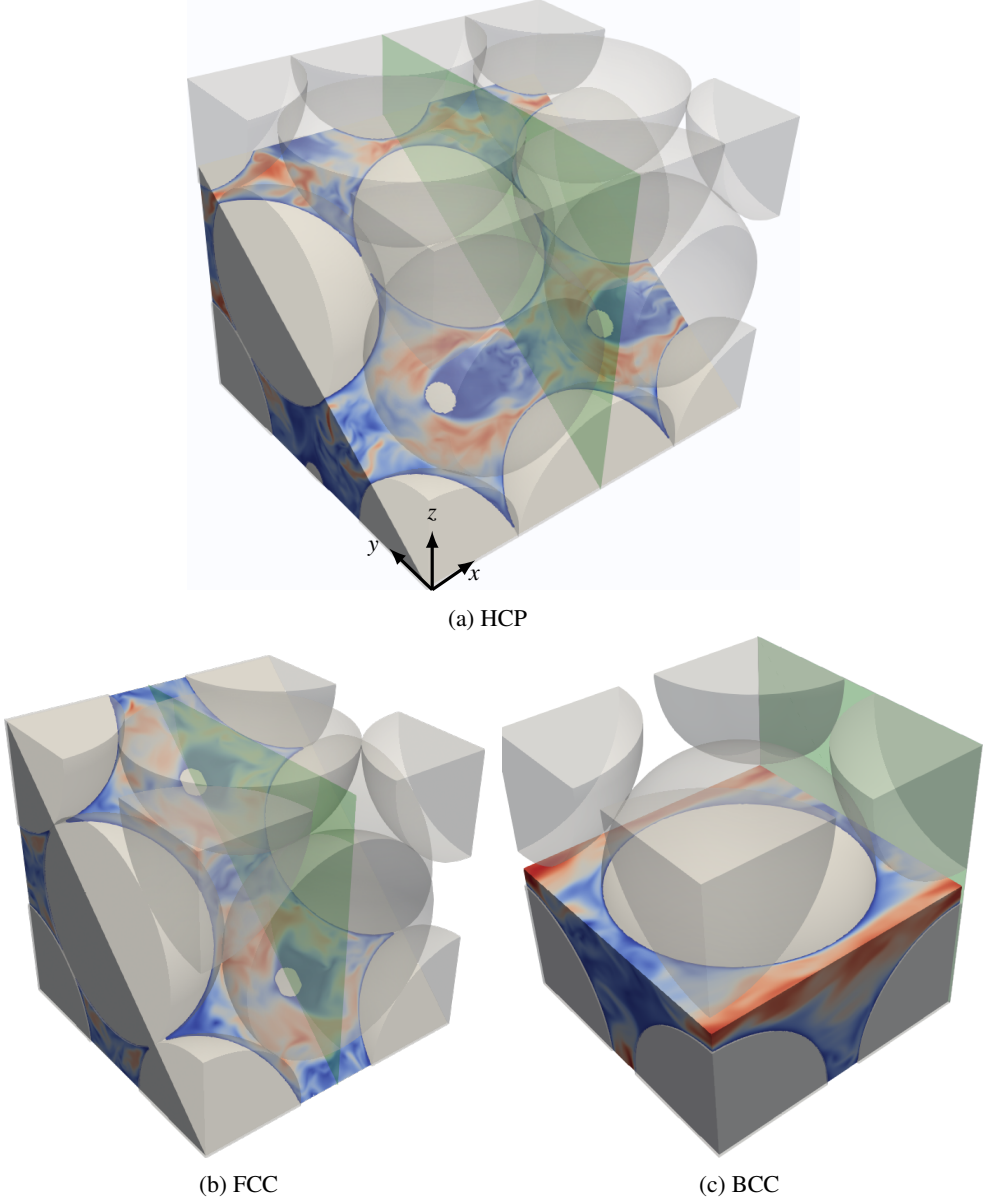
\begin{figure}
    \centering
    \subcaptionbox{HCP \label{sub-fig:hcp}}{
    \begin{tikzpicture}
        \node [above right,inner sep=0] (image) at (0,0) {
        \includegraphics[width=.6\linewidth]{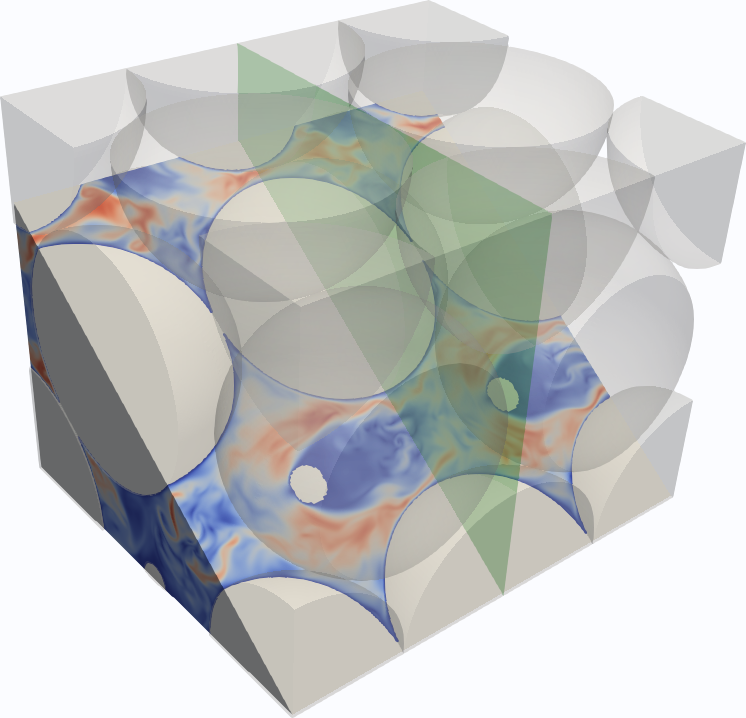}
        };

        \begin{scope}[shift={(image.south west)},
                      x={($(image.south east)-(image.south west)$)},
                      y={($(image.north west)-(image.south west)$)}
                     ]
        \draw[-latex,very thick,black] (0.4,0.0) -- ++(+0.07,0.05) node[pos=1,below]{$x$};
        \draw[-latex,very thick,black] (0.4,0.0) -- ++(-0.07,0.07) node[pos=1,left]{$y$};
        \draw[-latex,very thick,black] (0.4,0.0) -- ++(+0.0,0.1) node[pos=1,above]{$z$};
        \end{scope}
      \end{tikzpicture}
    }
    \subcaptionbox{FCC \label{sub-fig:fcc}}{
       \includegraphics[width=.47\linewidth]{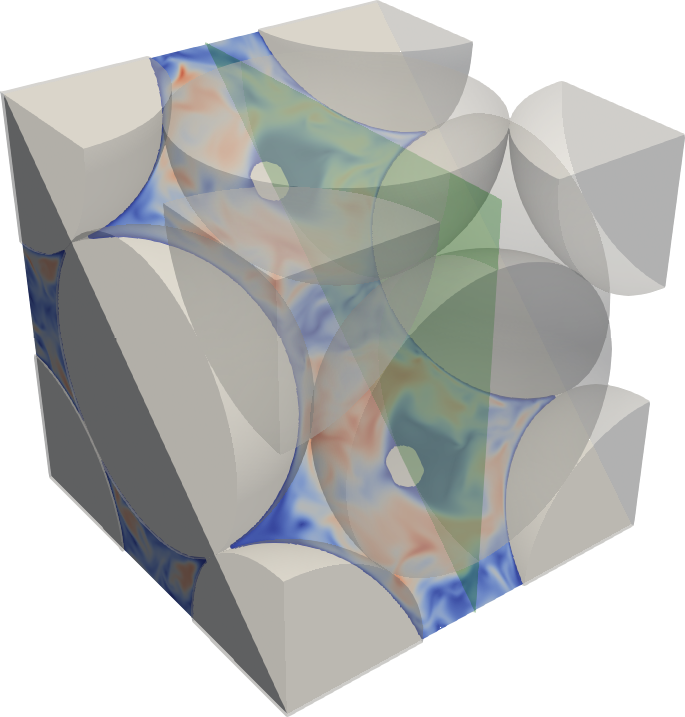}
    }
    \subcaptionbox{BCC\label{sub-fig:bcc}}{
       \includegraphics[width=.47\linewidth]{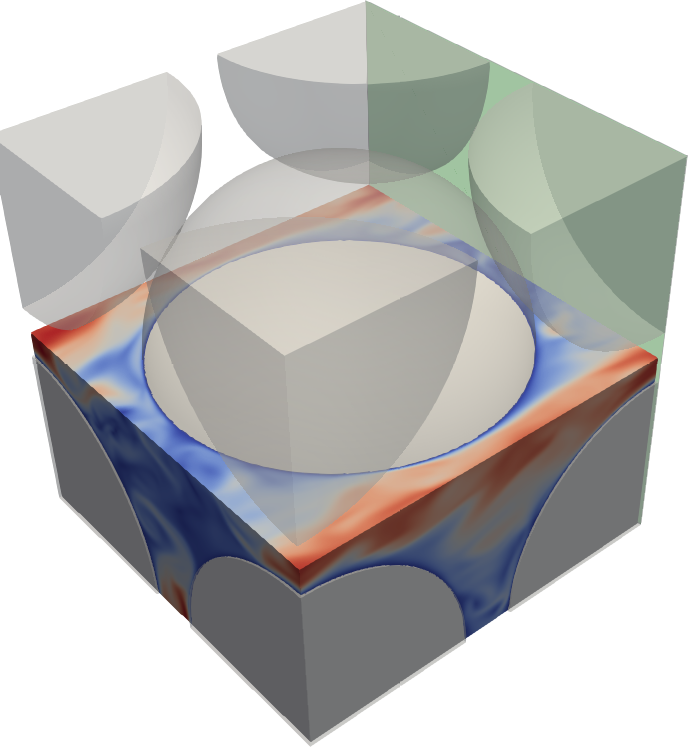}
    }
    \caption{Three sphere pack geometries being considered in
    this study. 
    To highlight the pore-scale flow features, the magnitude of an instantaneous velocity field is included, where red and blue colours indicate high and low magnitudes respectively.
    The cut planes correspond to a symmetry plane for each sphere pack: HCP, $\frac{\sqrt{3}}{3}y + \frac{\sqrt{6}}{3}z=0$; FCC, $y -z= 0$; BCC, $z=\frac{1}{2}d$.
    Green shaded planes correspond to the cross-stream planes for examining the pore-scale flow features:
    HCP, $x=d$; FCC, $x=\sqrt{2}d/2$; BCC, $x=2d/\sqrt{3}$
    }
    \label{fig:geometry}
\end{figure}

In this contribution we consider three different close-pack arrangements of uniform 
spheres, namely: hexagonal closed pack (HCP), face-centred cubic (FCC) and body-centred cubic (BCC) sphere packs \citep{conway_sphere_1999}, shown in figure \ref{fig:geometry}.
Those sphere packs are characterised by the sphere diameter 
$d$ and the porosity $\porosity$ (cf. table \ref{tab:geometric_properties}).

\begin{table}
    \centering
    \begin{tabular}{>{\raggedright}p{0.05\textwidth}>{\raggedright}p{0.08\textwidth}>{\raggedright}p{0.15\textwidth}>{\raggedright}p{0.22\textwidth}>{\raggedright}p{0.15\textwidth}p{0.25\textwidth}}
         & porosity $\porosity$ & permeability $\permeability/d^2$ & weighted pore-volume-to-surface ratio $\Lambda/d$ & static viscous tortuosity $\alpha_0$ & high-frequency limit of the dynamic tortuosity $\alpha_{\infty}$ \\
         \hline
         HCP & $0.259$ & $1.73\times 10^{-4}\,{}^{a}$ &  $5.9\times 10^{-2}\,{}^{c}$ & $2.66\,{}^{d}$ & $1.62\,{}^{c}$ \\
         FCC & $0.259$ & $1.73\times 10^{-4}$ & $6.2\times 10^{-2}\,{}^{b}$ & $2.65$ & $1.61\,{}^{b}$ \\
         BCC & $0.320$ & $5.16\times10^{-4}$ & $9.6\times 10^{-2}\,{}^{b}$ & $2.22$ & $1.47\,{}^{b}$ \\
         \hline
    \end{tabular}%
    
    
    \begin{minipage}{\textwidth}
        \raggedright
        $\,{}^{a}$ \citep{Sakai2020}\\
        $\,{}^{b}$ \citep{Chapman1992}\\
        $\,{}^{c}$ \citep{Unglehrt2023}\\
        $\,{}^{d}$ \citep{Zhu2016}
    \end{minipage}
    \caption{Geometric properties of the hexagonal close-packed, face-centred cubic and body-centred cubic sphere packs.}
    \label{tab:geometric_properties}
\end{table}

The triply-periodic boundary conditions are applied at each end of the numerical 
domains, which have the spatial extents of $[L_x, L_y, L_z]$ in the Cartesian
coordinate directions.
Note that whilst we simulate a single unit cell for the FCC and the BCC
configurations,
the domain extent of the HCP case is doubled in $x$-direction in order to
accommodate the symmetry breaking in the streamwise direction that appears
when the flow state is highly chaotic \citep{Sakai2020}.

In the present study, our accelerating porous media flow is characterised by the so-called \textit{Hagen number}: 

\begin{equation}
    \Hg = \frac{\left\vert\bnabla\!\intrinsicavg{p}\right\vert d^3}{\rho \nu^2} \ ,
\end{equation}
\noindent
whilst the
(statistically) steady-state Reynolds number ($Re_\mathrm{steady}$) is often used to describe the steady-state flow, which is simply a function of $\Hg$.
Note that by normalising with $\rho$, $d$ and $\left\vert\bnabla\!\intrinsicavg{p}\right\vert$,
the governing equations \eqref{eqn:nse-dimensional} can be expressed in the following dimensionless form such that the flow field depends solely on $\Hg$:

\begin{align}
    \hat{\bnabla} \bcdot \hat{\boldsymbol{u}} &= 0\\
    \hat{\partial_t} \hat{\boldsymbol{u}} + \hat{\bnabla} \bcdot (\hat{\boldsymbol{u}} \otimes \hat{\boldsymbol{u}}) &= 
    -\hat{\bnabla} \hat{p} + \frac{1}{\sqrt{\Hg}} \hat{\Delta} \hat{\boldsymbol{u}} + \boldsymbol{e}_x\ ,
    \label{eqn:nse}
\end{align}

\noindent
where $\hat{(*)}$ represents non-dimensionalised variables and operators.
The governing parameter $\Hg$ can be seen as the ratio of the intrinsic pressure gradient to the viscous forces, and have been referred to as a pressure-gradient-based Reynolds number in other works \citep[e.g.][p.~416]{Schlichting.2017}.
In this study, initially-resting fluid is accelerated by a constant intrinsic 
pressure gradient until a steady-state is established.

To ensure the comparability to the existing steady flow studies, however, we define the aforementioned Reynolds number as following.
This Reynolds number is based on the superficial (volume-averaged) streamwise velocity:

\begin{equation}
    \superficialavg{u} = \frac{1}{V}\int_{V_\mathrm{f}} u(\Vec{x},t)\,\mathrm{d}V\ ,
\end{equation}

\noindent
where $V_{\mathrm{f}}$ is the fluid volume, whilst $V = L_x\,L_y\,L_z$ includes the volume of spheres as well in which the velocity is set at zero.
Note that the corresponding \textit{intrinsic} (fluid-volume-averaged) velocity
can be determined by using porosity as $\intrinsicavg{u} = \superficialavg{u}/\porosity$.
The corresponding Reynolds number is defined as:

\begin{equation}
    \Rey_\steady = \frac{\overline{\superficialavg{u}}d}{\nu} \ ,
\end{equation}

\noindent
where $\overline{(*)}$ represents (statistically) steady-state.
Alternatively, the Reynolds number definition based on the hydraulic diameter
($\Rey_\mathrm{H}$) is also commonly used, which is particularly useful to compare porous media with different
porosity as in this study.
This Reynolds number is related to $\Rey_\steady$ in the following manner \citep{Wood2020}: 

\begin{equation}
    \Rey_\mathrm{H} = \frac{\Rey_\steady}{1-\porosity}\ .
\end{equation}

Finally, a set of volume-averaged quantities, such as the superficial velocity and energy, were sampled at least every 40 timesteps from the start-up \textit{on-the-fly} for each simulation, in addition the velocity and pressure distributions on a cross-flow plane at the full spatial resolution.
Additionally, a series of snapshots of the complete flow field were saved for some key simulation cases.

\subsection{Hexagonal close-packing (HCP)}
\label{sub-sec:dataset_HCP}

The results of this high-fidelity DNS dataset consisting of 16 cases were first published in \cite{Sakai2020}.
For this study, its subset of 11 cases are evaluated (cf. table \ref{tab:hcp-param}).
The porosity of this HCP sphere pack is at the close-pack limit of $\epsilon = 1-\frac{\pi}{3\sqrt{2}} \approx 0.259$,
which means all the spheres are in contact with their twelve neighbours.
The spatial extents of the numerical domain are: $[L_x, L_y, L_z]/d = [2, \sqrt{3}, \frac{2\sqrt{6}}{3}]$.   
This sphere pack geometry with respect to the considered primary flow direction
features a split-and-merging pattern of alternating smaller and larger pores, 
which is highlighted in figure \ref{sub-fig:hcp} by means of a symmetry-plane cutout.
Due to this geometric constraint, the flow is accelerated significantly through the
smaller pores, before it reduces the velocity magnitude in the subsequent 
larger cavities.
Once $\Hg$ becomes sufficiently high, the primary flow starts to separate
from the sphere surface behind the sphere contact points inside the larger pores 
(cf. \citet[][]{Sakai2020}, also the velocity magnitude distribution in figure \ref{sub-fig:hcp}).

The $\Hg$ range being covered by this dataset is 
$3.25 \times 10^{-3} \leq \Hg \leq 1.3 \times 10^{7}$, spanning over linear, steady nonlinear,
unsteady nonlinear and turbulent flow regimes.
We employed equidistant grid point distributions in $x$, $y$, $z$-directions, with
$160$ finite-volume cells per diameter up to $\Hg=4.9\times 10^{5}$, and $320$ points per diameter
between $6.5 \times 10^{5} \leq \Hg \leq 1.3 \times 10^{7}$.
The corresponding grid convergence study was carried out and presented in \citet[][Fig. 3]{Sakai2020} and \citet[][Figure 2]{Unglehrt2022}.
This highly adequate grid resolution was adopted to resolve the very thin 
potential-flow type boundary layers forming on the sphere surface when the flow
is suddenly accelerated from the rest.
For adequate temporal accuracy, the maximum CFL number was kept below 0.16.

\begin{table}
    \centering
    \begin{tabular}{c c c c c c c}
         CASE & CASE in SM20 & $\Hg$ & $\Rey_\steady$ & $\Rey_\mathrm{H}$ & Colour/marker & $d/\Delta x$ \\
         \hline
         H1 & L6   & $6.5\times 10^{3}$ & $1$   & $1.35$ & \textcolor{nintendo_yellow}{+} & $160$ \\
         H2 & SNL1 & $6.5\times 10^{4}$ & $10$  & $13.5 $ & \textcolor{nintendo_red}{+} & $160$ \\
         H3 & SNL2 & $3.3\times 10^{5}$ & $36$  & $48.6 $ & \textcolor{nintendo_red}{$\medcirc$} & $160$ \\
         H4 & SNL3 & $4.9\times 10^{5}$ & $48$  & $64.9 $ & \textcolor{nintendo_red}{$\square$} & $160$ \\
         H5 & SNL4 & $6.5\times 10^{5}$ & $59$  & $79.7$ & \textcolor{nintendo_red}{$\bigtriangleup$} & $320$ \\
         H6 & UNL1 & $1.3\times 10^{6}$ & $91$  & $123$ & \textcolor{nintendo_green}{+} & $320$ \\
         H7 & UNL2 & $2.6\times 10^{6}$ & $138$ & $186$ & \textcolor{nintendo_green}{$\medcirc$} & $320$ \\
         H8 & T1   & $5.2\times 10^{6}$ & $209$ & $282$ & \textcolor{nintendo_blue}{+} & $320$ \\
         H9 & T2   & $7.8\times 10^{6}$ & $254$ & $343$ & \textcolor{nintendo_blue}{$\medcirc$} & $320$ \\
         H10 & T3   & $1.0\times 10^{7}$ & $305$ & $412$ & \textcolor{nintendo_blue}{$\square$} & $320$ \\
         H11 & T4   & $1.3\times 10^{7}$ & $347$ & $469$ & \textcolor{nintendo_blue}{$\bigtriangleup$} & $320$ \\
         \hline
    \end{tabular}
    \caption{Numerical and physical parameters of the HCP simulations examined in this study.
    The entries on the second column correspond to the case names introduced in \citet{Sakai2020} and referred in \citet{Unglehrt2023}, whose suffixes describe the flow state, as: L, linear; SNL, steady nonlinear; UNL, unsteady nonlinear; T, turbulent}
    \label{tab:hcp-param}
\end{table}

\subsection{Face-centred cubic packing (FCC)}
\label{sub-sec:dataset_FCC}


This dataset of the flow though the FCC sphere pack consists of 11 DNS results, which covers
$5.6 \times 10^{1} \leq \Hg \leq 1.2 \times 10^7$ (cf. table \ref{tab:fcc-bcc-param}).
The spatial extents of the numerical domain are $[L_x, L_y, L_z]/d = [\sqrt{2}, \sqrt{2}, \sqrt{2}]$.
The porosity of the sphere pack is at the close-pack limit of $\epsilon \approx 0.259$,
which is identical to that of the HCP configuration.
The  split-and-merging pattern mentioned for HCP is also featured in 
this geometry, meaning the fluid inside one larger pore flows into subsequent 
four smaller pores before merging in the following larger pore.
Previously, it was reported that a small wake-like reverse flow region forms directly 
behind one of the spheres enclosing the larger pore space \citep{HILL2001a,He2019}, 
indicating the existence of significant flow separations under a strong inertial influence.
Such geometric and flow feature are visualised in figure \ref{sub-fig:fcc}.

We employed equidistant grid point distributions in $x$, $y$, $z$-directions, with
$158$ finite-volume cells per diameter in the cases up to $\Hg = 2.3 \times 10^6$, and 
$316$ cells per diameter between $\Hg = 4.6 \times 10^{6}$ and $1.2 \times 10^{7}$.
Note that this grid resolution is significantly higher than the required resolution reported in \cite{He2019} to achieve a grid-convergence of the mean flow as well as
second-order statistics in their FCC porous media flow up to $\Rey_\steady=370$.
For adequate temporal accuracy, the maximum CFL number was kept below 0.31.
%
%
The simulation results were validated against the data available in literature, and are summarised in appendix \ref{app:dnsdata_fcc}.
Moreover, the flow state corresponding to each case was determined based on the geometric symmetries of the pore-scale flow, and also summarised in appendix \ref{app:dnsdata_fcc}.
Consequently, we categorise that F1 belongs the linear flow regime, the cases from F2 and F5 are in the steady nonlinear regime, F6 belongs to the unsteady nonlinear flow, and F7--11 fall into the turbulent/chaotic flow category.

\begin{table}
    \centering
    \begin{tabular}{c c c c c c c}
         CASE & Flow state & $\Hg$ & $\Rey_\steady$ & $\Rey_\mathrm{H}$ & Colour/marker & $d/\Delta x$ \\
         \hline
         F1 & L & $5.6\times 10^{1}$ & $9.7 \times 10^{-3}$   & $1.3 \times 10^{-2}$ & \textcolor{nintendo_yellow}{+} & $158$ \\
         F2 & SNL & $5.8\times 10^{4}$ & $9.6$  & $13 $ & \textcolor{nintendo_red}{+} & $158$ \\
         F3 & SNL & $2.9\times 10^{5}$ & $37$  & $49.8 $ & \textcolor{nintendo_red}{$\medcirc$} & $158$ \\
         F4 & SNL & $4.3\times 10^{5}$ & $49$  & $66.5 $ & \textcolor{nintendo_red}{$\square$} & $158$ \\
         F5 & SNL & $5.8\times 10^{5}$ & $60$  & $80.9$ & \textcolor{nintendo_red}{$\bigtriangleup$} & $158$ \\
         F6 & UNL & $1.2\times 10^{6}$ & $94$  & $127$ & \textcolor{nintendo_green}{+} & $158$ \\
         F7 & T & $2.3\times 10^{6}$ & $146$ & $197$ & \textcolor{nintendo_blue}{+} & $158$ \\
         F8 & T & $4.6\times 10^{6}$ & $223$ & $301$ & \textcolor{nintendo_blue}{$\medcirc$} & $316$ \\
         F9 & T & $6.9\times 10^{6}$ & $279$ & $376$ & \textcolor{nintendo_blue}{$\square$} & $316$ \\
         F10 & T & $9.2\times 10^{6}$ & $332$ & $449$ & \textcolor{nintendo_blue}{$\bigtriangleup$} & $316$ \\
         F11 & T & $1.2\times 10^{7}$ & $378$ & $511$ & \textcolor{nintendo_blue}{$\bigtriangledown$} & $316$ \\
         \hline
         B1 & L & $2\times 10^{2}$ & $0.1$   & $1.5 \times 10^{-1}$ & \textcolor{nintendo_yellow}{+} & $166$ \\
         B2 & SNL & $2\times 10^{4}$ & $9.6$  & $14 $ & \textcolor{nintendo_red}{+} & $166$ \\
         B3 & SNL & $1\times 10^{5}$ & $40$  & $59$ & \textcolor{nintendo_red}{$\medcirc$} & $166$ \\
         B4 & SNL & $1.5\times 10^{5}$ & $56$  & $82$ & \textcolor{nintendo_red}{$\square$} & $166$ \\
         B5 & SNL & $2\times 10^{5}$ & $70$  & $103$ & \textcolor{nintendo_red}{$\bigtriangleup$} & $166$ \\
         B6 & SNL & $4\times 10^{5}$ & $122$  & $165$ & \textcolor{nintendo_red}{$\bigtriangledown$} & $166$ \\
         B7 & UNL & $8\times 10^{5}$ & $166$ & $244$ & \textcolor{nintendo_green}{+} & $166$ \\
         B8 & T & $1.6\times 10^{6}$ & $242$ & $356$ & \textcolor{nintendo_blue}{$+$} & $332$ \\
         B9 & T & $2.4\times 10^{6}$ & $314$ & $462$ & \textcolor{nintendo_blue}{$\medcirc$} & $332$ \\
         B10 & T & $3.2\times 10^{6}$ & $410$ & $604$ & \textcolor{nintendo_blue}{$\square$} & $332$ \\
         B11 & T & $4\times 10^{6}$ & $429$ & $631$ & \textcolor{nintendo_blue}{$\bigtriangleup$} & $332$ \\
         \hline
    \end{tabular}
    \caption{Numerical and physical parameters of the FCC (F cases) and the BCC (B cases) simulations.
    The entries on the second column correspond to the flow state, as: L, linear; SNL, steady nonlinear; UNL, unsteady nonlinear; T, turbulent
    }
    \label{tab:fcc-bcc-param}
\end{table}

\subsection{Body-centred cubic packing (BCC)}
\label{sub-sec:dataset_BCC}

This dataset of the flow through a BCC sphere pack consists of 11 DNS results
(cf. table \ref{tab:fcc-bcc-param}).
The range of $\Hg$ covered by this dataset is 
$2 \times 10^{2} \leq \Hg \leq 4 \times 10^{6}$.
The spatial extents of the numerical domain are: $[L_x, L_y, L_z]/d = [\frac{2}{\sqrt{3}}, \frac{2}{\sqrt{3}}, \frac{2}{\sqrt{3}}]$.
The porosity of this geometry is at $\epsilon \approx 0.32$,
which is approximately $23\%$ higher than the other two configurations.
This higher porosity is readily visible in figure \ref{sub-fig:bcc}, where large,
straight unobstructed pore channels pointing in the streamwise direction are featured on four sides (i.e. top, bottom, left and right) of the central sphere.
A grid resolution of $166$ finite-volume cells per diameter is adopted for the cases up to
$\Hg = 8 \times 10^5$, whereas $332$ cells per diameter resolution is used for the cases 
between $\Hg = 1.6 \times 10^6$ and $4 \times 10^6$.
The maximum CFL number was kept below 0.35.
As for FCC, the simulation results were compared to the data available in literature, and are summarised in appendix \ref{app:dnsdata_bcc}.

As for HCP and FCC, the flow state corresponding to each case was determined based on the geometric symmetries of the pore-scale flow as summarised in appendix \ref{app:dnsdata_bcc}.
Accordingly, we categorise B1 to be in the linear flow regime; the cases B2 through B6 are in the steady nonlinear flow regime; B7 is in the unsteady nonlinear regime; and the cases B8--11 to be in the turbulent/chaotic flow regime.

\section{Time scales in transient flow}
\label{sec:time-scale}

Here, we discuss the evolution of the start-up flow with a special focus on the governing time scales of different physical processes.
First, we perform a dimensional analysis to identify two time scales that are relevant in the accelerating flow through porous media.
Then, based on the identified time scales, we perform an \textit{a-priori} order-of-magnitude analysis, leading to a hypothesis that one of the two time scales controls the onset of the nonlinearity emerging in the pore space. 
Finally, the hypothesis is tested against the DNS data.

\subsection{Dimensional analysis}
\label{sub-sec:da}

We assume that the following fundamental parameters describe the start-up flow field within the porous media $\Vec{u}(\Vec{x},t)$: The fluid density $\rho$ and kinematic viscosity $\nu$, the sphere diameter $d$, the intrinsic pressure gradient driving the flow $\left\vert\bnabla\!\intrinsicavg{p}\right\vert$, and the elapsed time $t$.
Porosity $\porosity$ and the permeability $\permeability$ of the porous media are left out, since they are pure functions of the geometry and remained constant for each sphere pack.

Depending on our choice of the repeating variables, different dimensionless groups can be formulated, as well as characteristic time and velocity scales. 
For instance, with $d$, $\left\vert\bnabla\!\intrinsicavg{p}\right\vert$ and $\nu$, the following functional relation of the corresponding dimensionless groups can be formed:
\begin{equation}
\frac{u\, d}{\nu} = \Rey = \frac{u\, \tauvisc}{d} = \Phi\!\left(\Hg,\; \frac{t}{\tauvisc}  \right)\quad , 
\end{equation}
\noindent
where $\Phi$ is an arbitrary function, whereas the normalisation factor $\tauvisc$ is a \textit{viscous} time scale:
\begin{equation}
    \tauvisc=\frac{d^2}{\nu}\quad .
\end{equation}
This time scale controls the flow evolution when it is predominantly influenced by viscosity $\nu$ ---one of the repeating variables--- and represents the required time to reach an equilibrium by molecular diffusion.
Moreover, since $\permeability$ and $\porosity$ are constant the aforementioned energy time scale $\tauen$ can be expressed in terms of $\tauvisc$, as:
\begin{equation}
      \tauen=\frac{\alpha_0}{\porosity} \left( \frac{\permeability}{d^2} \right) \tauvisc \quad ,
      \label{eqn:tauen}
\end{equation}
\noindent
where $\alpha_0$ is static viscous tortuosity, which is a purely geometrical-hydrodynamic property of the porous structure (cf. table \ref{tab:geometric_properties}).
This tortuosity $\alpha_0$ is the low-frequency ratio by which the actual kinetic energy of fluid in a porous medium exceeds the kinetic energy computed from the superficial velocity, due to the non-uniform pore-scale velocity field created by viscous flow \citep{Zhu2014}.

On the other hand, if we select $\rho$ instead of $\nu$ as a repeating variable ---corresponding to the situations where the viscosity effect is not predominant--- the following functional relation can be found:

\begin{equation}
  \frac{u\, \tauinv}{d} = \Psi\!\left(\Hg,\; \frac{t}{\tauinv}  \right)\quad ,
  \label{eqn:da-inv}
\end{equation}
where $\Psi$ is another arbitrary function, whilst $\tauinv$ is an \textit{inviscid} time scale:
\begin{equation}
    \tauinv=\left(\frac{\rho d}{\left\vert\bnabla\!\intrinsicavg{p}\right\vert}\right)^{\frac{1}{2}}.
\end{equation}
The mechanistic implication of the inviscid time scale at the pore scale will be analysed in the following section.
Moreover, the square of the ratio of the viscous to the inviscid time scale forms the aforementioned Hagen number, viz.:
   \begin{equation}
       \left(\frac{\tauvisc}{\tauinv}\right)^2 = \Hg \ ,
       \label{eqn:hagen}
   \end{equation}
which can also be seen as the ratio of the intrinsic pressure gradient to the viscous forces.
Notice that $\Hg$ emerges as a governing dimensionless parameter independent of the repeating variable choices,
whilst $\Rey$ and $\Rey_{\mathrm{steady}}$ are unique functions of $\Hg$ and dimensionless time.
\subsection{Onset of nonlinearity in the boundary layer}
\label{sub-sec:onset}

\begin{figure}
    \centering
    \includegraphics[width=0.7\linewidth]{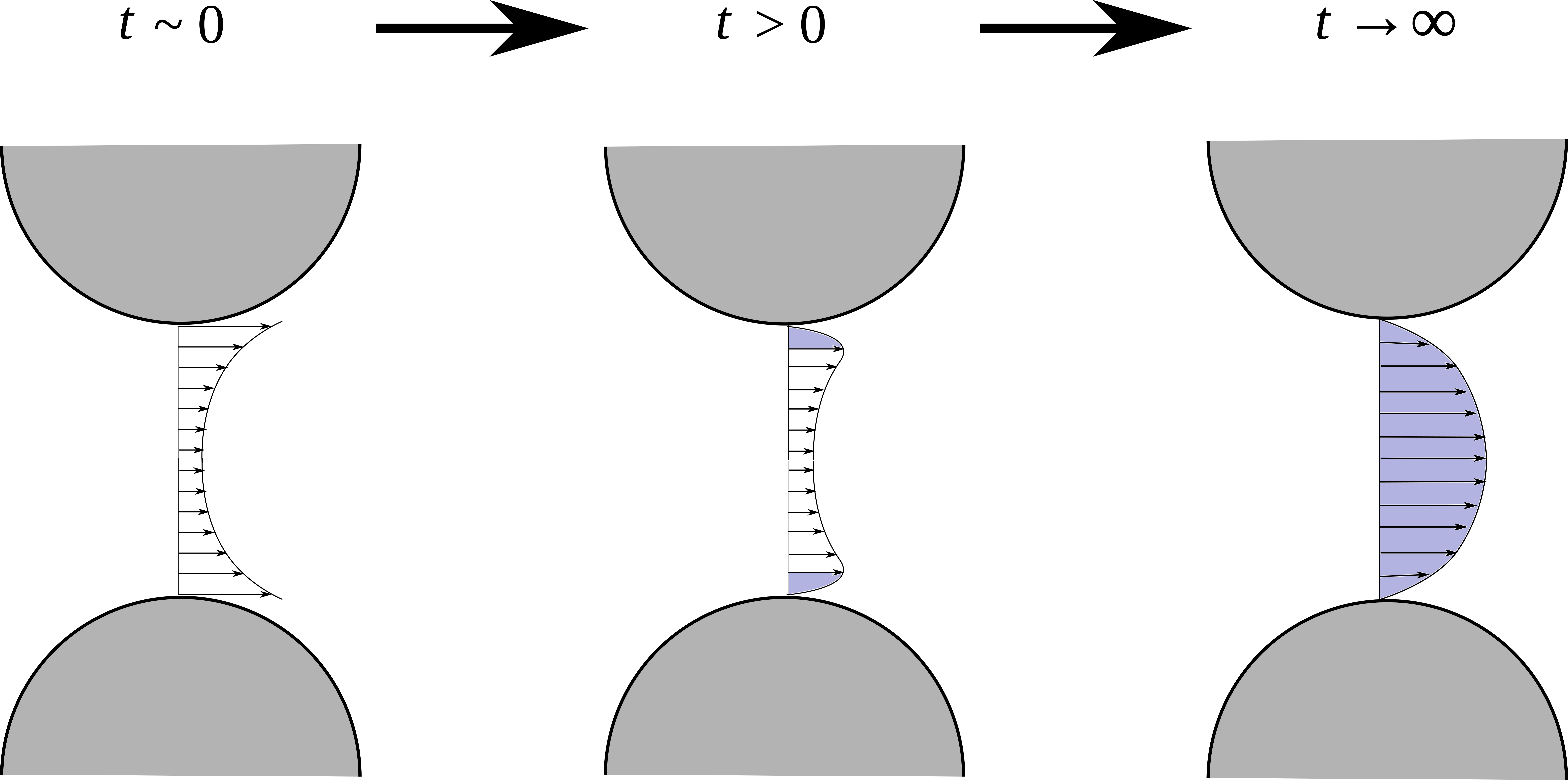}
    \caption{Schematic of linear transient flow development from a potential flow at $t=0$ over a potential core flow with growing Stokes boundary layers (blue shaded regions) for small $t>0$ towards a fully developed Stokes flow for $t\to \infty$.}
    \label{fig:potential_bl}
\end{figure}

Next, we analyse the Stokes boundary layer developing on the sphere surfaces in the flow through the porous medium.
As discussed in \S \ref{sec:intro}, such boundary layer is initially infinitely thin and can be found beneath the irrotational outer flow (see also figure \ref{fig:potential_bl} for a linear flow example).
Initially, the outer flow behaves as an ideal potential flow due to inexistence of actual flow resistance, and accelerates at a constant rate in response to a constant intrinsic pressure gradient, whilst the boundary layer thickens over time due to viscous diffusion.
Since the nonlinearity of the flow generally arises from the convective term in the governing equations, it is initially at zero within the boundary layer and builds up with the flow velocity.

In this analysis, we perform an order-of-magnitude estimation of the terms in the unsteady laminar boundary layer equations \citep{Schlichting.2017}, with a focus on high-$\Hg$ flows such that $\tauinv \ll \tauvisc$ (cf. equation \ref{eqn:hagen}).
Under this assumption and a help of characteristic scales, the boundary layer equations can be approximated as:
\begin{equation}
       \underbrace{O\left(\frac{U}{T}\right)}_{\text{local acceleration}} 
       +\underbrace{O\left(\frac{U^2}{L}\right)}_{\text{convection}}
       = 
       \underbrace{O\left(\frac{1}{\rho} \frac{P}{L} \right)}_{\text{pressure gradient}}
       + \underbrace{O\left(\nu \frac{U}{\delta^2} \right)}_{\text{viscous diffusion}}
       + \underbrace{O\left(\frac{\left\vert\bnabla\!\intrinsicavg{p}\right\vert}{\rho}\right)}_{\text{driving force}}\, .
\end{equation}
\noindent
During the initial acceleration, the characteristic time $T$ can be equated to the elapsed time $t$ since the start (i.e. $T \sim t$).
On the other hand, the characteristic velocity scale can be estimated from the accelerating potential flow solution as $U \sim \rho^{-1} |\bnabla\!\intrinsicavg{p}|\, t$, and the pressure scale is determined by the macroscopic pressure gradient ($P \sim |\bnabla\!\intrinsicavg{p}|\,d$). 
Notice that the boundary layer thickness is taken proportional to the diffusion length ($\delta \sim \sqrt{\nu t}$), whereas the tangential length scale in the other terms is taken as the sphere diameter ($L \sim d$).
The former estimation is deduced from the high-$\Hg$ flow assumption, where the viscous diffusion of momentum is confined within very thin boundary layers on the sphere surfaces. 
By inserting these estimates and dividing by $\rho^{-1} |\bnabla\!\intrinsicavg{p}|$, we obtain the dimensionless order of magnitude estimation, as:
\begin{equation}
	\underbrace{1}_{\text{local acceleration}} 
    + \underbrace{\left(\frac{t}{\tauinv}\right)^2}_{\text{convection}}
    = 
    \underbrace{1}_{\text{pressure gradient}} 
    + \underbrace{1}_{\text{viscous diffusion}} 
    + \underbrace{1}_{\text{driving force}} \,.
\end{equation}

\noindent
Consequently, the local acceleration, the pressure gradient and the diffusive term have the same order of magnitude, whereas the importance of the nonlinear convective term, which is insignificant initially, grows with time.
At $t \sim \tauinv$, the nonlinear term has the same order-of-magnitude as the other terms in the boundary layer equations, signalling the onset of nonlinearity.

Conversely, in case of low-$\Hg$ flows, the boundary layer fills entire pore space almost immediately since $\tauinv \sim \tauvisc$.
Consequently, the velocity scale $U$ can no longer be represented by a potential flow solution, but rather be the steady-state velocity according to Darcy's law (i.e. $U \sim d^2 \rho^{-1} \nu^{-1} |\bnabla\!\intrinsicavg{p}|$). The length scale for the viscous diffusion term is $\delta \sim d$.
Then we obtain:
\begin{equation}
	\underbrace{\left(\frac{\tauvisc}{t}\right)}_{\text{local acceleration}}
    + \underbrace{\Hg}_{\text{convection}} 
    = 
    \underbrace{1}_{\text{pressure gradient}} 
    + \underbrace{1}_{\text{viscous diffusion}} 
    + \underbrace{1}_{\text{driving force}} \,.
\end{equation}

\noindent
As expected, the local acceleration term has become time-dependent and diminishes at the large-time limit ($t \rightarrow \infty$).
The timing of the saturation is controlled by $\tauvisc$.
Meanwhile, the nonlinearity which is represented by the convective term is capped by the low-$\Hg$ assumption. 

\subsection{Scaling of the superficial velocity}
\label{sub-sec:nonlinear}

So far, we have shown that two distinct time scales (i.e. $\tauvisc$ or $\tauen$ and $\tauinv$) can be deduced from a dimensional analysis as the candidates for the governing time scales of the start-up flow through sphere packs.
On one hand, it is known that the viscous time scale controls the developing linear flow \citep{HILL2001,Zhu2014}.
On the other hand, we have shown that the onset of nonlinearity inside the pore-scale boundary layers should be timed by $\tauinv$ for accelerating nonlinear flows with high Hagen numbers (cf. \S \ref{sub-sec:onset}).
More specifically, the nonlinear inertial effect should become significant at $t \sim \tauinv$.
In the following, we test the above hypothesis by means of the DNS datasets.

\begin{figure}
    \centering
    \subcaptionbox{\label{sub-fig:ubulk-time-hcp-visc}}{
      \begin{tikzpicture}
        \node [above right,inner sep=0] (image) at (0,0) {
        \includegraphics[width=0.47\textwidth]{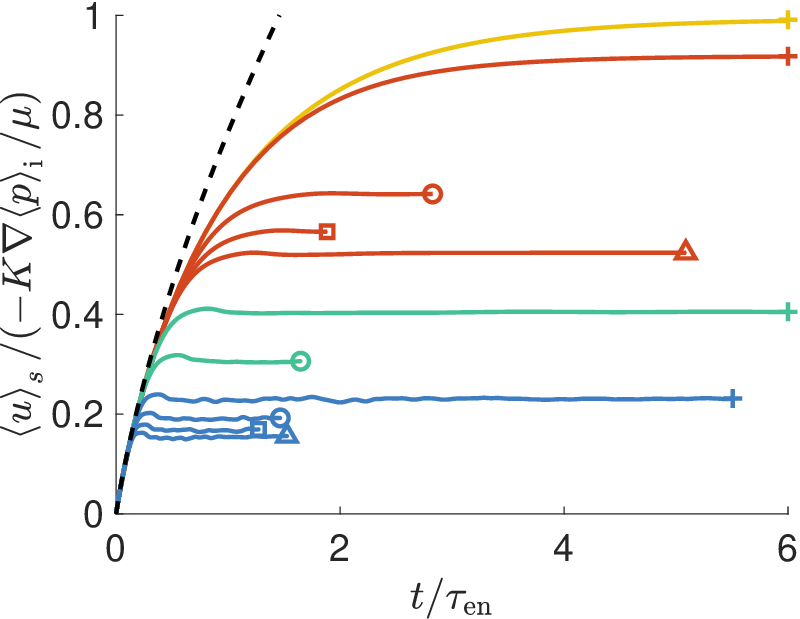}
        };

        \begin{scope}[shift={(image.south west)},
                      x={($(image.south east)-(image.south west)$)},
                      y={($(image.north west)-(image.south west)$)}
                     ]
        \draw[latex-,very thick,black] (0.45,0.25) -- ++(+0.2,+0.55) node[right,black]{$\Hg$};
        \end{scope}
      \end{tikzpicture}
      }
    \subcaptionbox{\label{sub-fig:ubulk-time-hcp-inv}}{
        \begin{tikzpicture}
        \node [above right,inner sep=0] (image) at (0,0) {
        \includegraphics[width=0.47\textwidth]{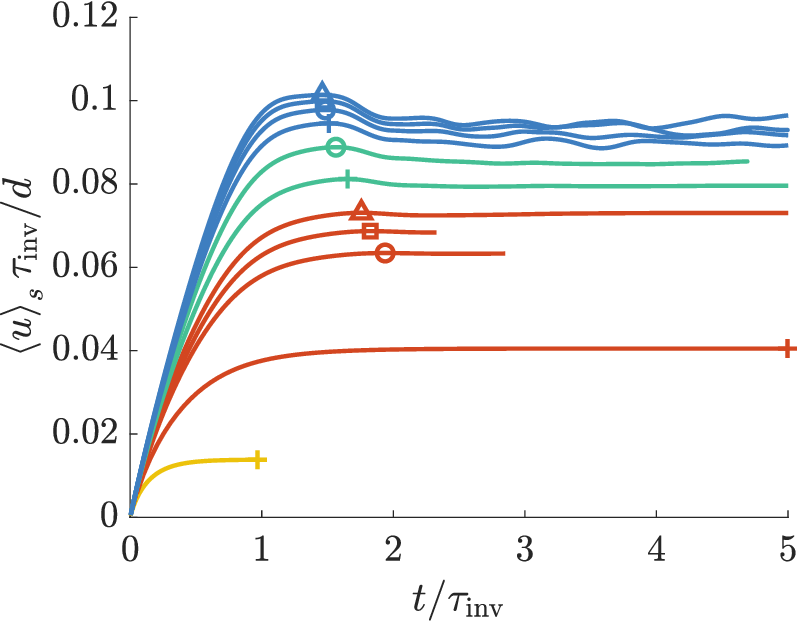}
        };

        \begin{scope}[shift={(image.south west)},
                      x={($(image.south east)-(image.south west)$)},
                      y={($(image.north west)-(image.south west)$)}
                     ]
        \draw[latex-,very thick,black] (0.65,0.9) -- ++(+0.0,-0.6) node[right,black]{$\Hg$};
        \end{scope}
      \end{tikzpicture}
      }
    \subcaptionbox{\label{sub-fig:ubulk-time-fcc-visc}}{
    \includegraphics[width=0.47\textwidth]{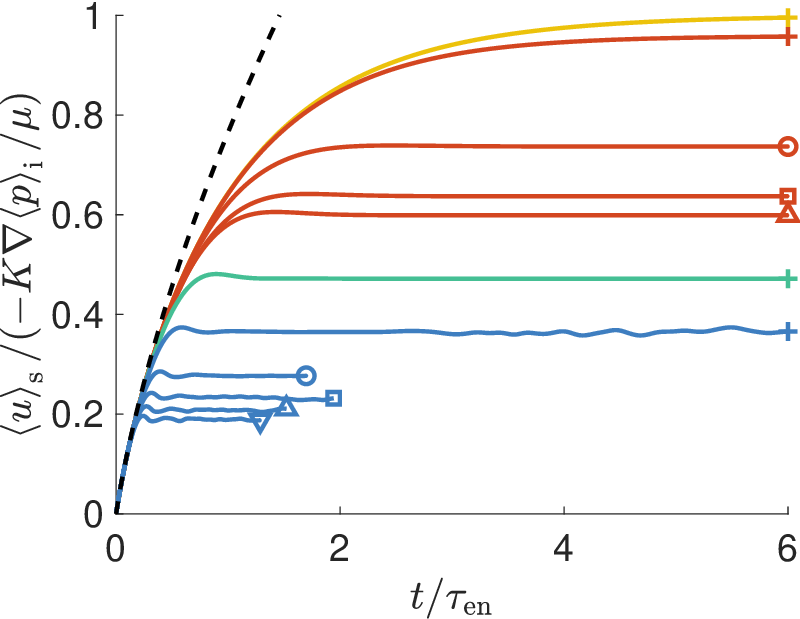}
    }
    \subcaptionbox{\label{sub-fig:ubulk-time-fcc-inv}}{
    \includegraphics[width=0.47\textwidth]{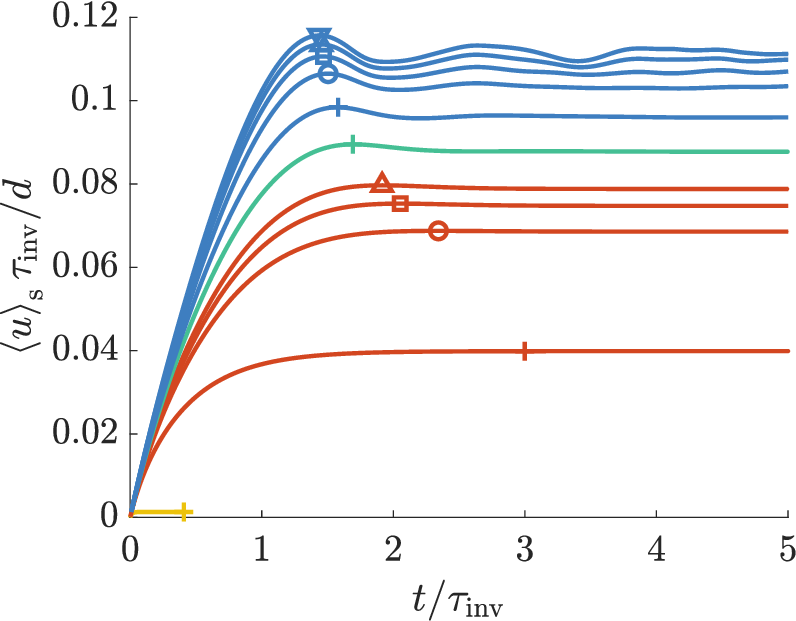}
    }
    \subcaptionbox{\label{sub-fig:ubulk-time-bcc-visc}}{
    \includegraphics[width=0.47\textwidth]{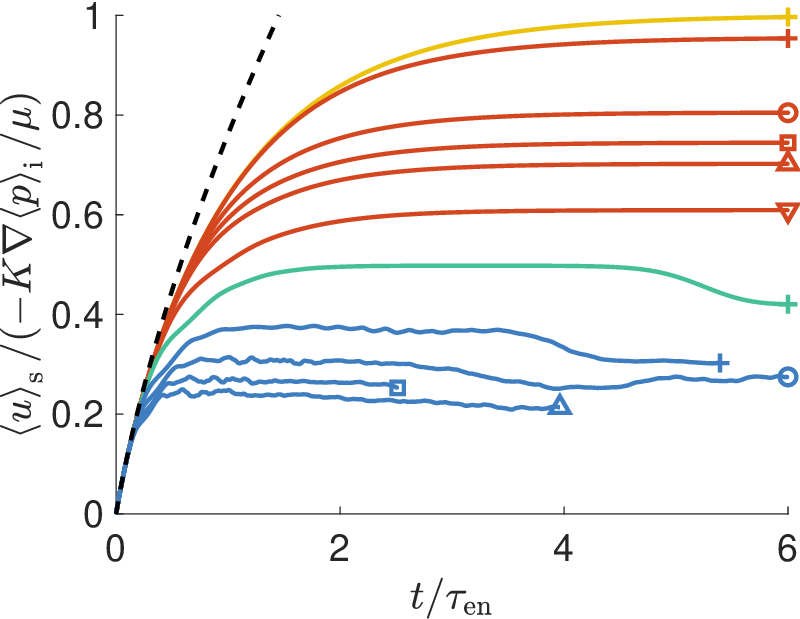}
    }
    \subcaptionbox{\label{sub-fig:ubulk-time-bcc-inv}}{
    \includegraphics[width=0.47\textwidth]{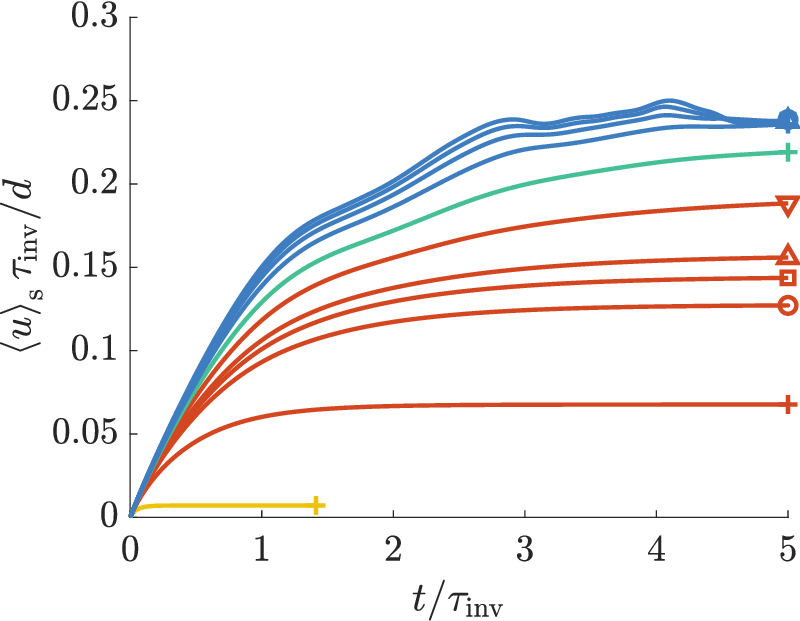}
    }
  \caption{
  Development of streamwise superficial velocity normalised by: (a,c,e) viscous scales; (b,d,f) inviscid scales.
  (a,b) HCP; (c,d) FCC; (e,f) BCC.
  Line colours and markers are according to tables \ref{tab:hcp-param} and \ref{tab:fcc-bcc-param}. 
  Black dashed lines are the solutions of equation \eqref{eqn:small-time}.
  }
  \label{fig:ubulk-time}
\end{figure}

We start by examining the temporal development of the superficial velocity $\superficialavg{u}$ through the HCP sphere pack shown in figure \ref{sub-fig:ubulk-time-hcp-visc}. 
In this figure, time is normalised by $\tauen$, which is proportional to $\tauvisc$ \eqref{eqn:tauen}.
Under this viscous normalisation, the velocity magnitude is normalised by the level predicted by Darcy's law.
Firstly, we observe for $\Hg \geq 3.3\times 10^5$ ($\Rey_{\mathrm{H}} \geq 48.6$, Case H3--11) that the steady-state velocity amplitude is significantly smaller than the Darcy's law level. 
The deviation from the linear relation grows continuously as $\Hg$ increases, and this trend is also present in the FCC and the BCC datasets (cf. figure \ref{sub-fig:ubulk-time-fcc-visc} and \ref{sub-fig:ubulk-time-bcc-visc}).
This $\Hg$-dependence is commonly known, and due to the emergence of the nonlinear advective drag \citep{Unglehrt2023}, which is usually accounted for by the Forchheimer term in steady flow \citep{Forchheimer1901,Whitaker:1996}.

Secondly, all curves, including both the linear and the nonlinear cases, become similar for a short time after the start-up ($t=0$) under this viscous normalisation. 
Moreover, the curves from all three sphere packs trace almost exactly the solutions of the following small-time asymptotics (cf. \citealp{Johnson.1987}, also \citealp[\S B.2.]{Unglehrt2023}):
\begin{equation}
    \rho \frac{\mathrm{d}\!\superficialavg{\Vec{u}}}{\mathrm{d}t} = 
    \underbrace{-\rho \sqrt{\nu} \frac{2}{\Lambda} \int_{0}^{t} \frac{\mathrm{d}\!\superficialavg{\Vec{u}}}{\mathrm{d}\tau} \frac{1}{\sqrt{\pi(t-\tau)}}\,\mathrm{d}\tau}_{\text{viscous friction \& pressure drag}}
    \underbrace{-\frac{\porosity}{\alpha_\infty}\nabla\!\intrinsicavg{p}}_{
    \text{effective driving force}}
    \ ,
    \label{eqn:small-time}
\end{equation}
where high-frequency limit of the dynamic tortuosity $\alpha_\infty$ and a weighted pore-volume-to-surface ratio $\Lambda$ are geometric properties and defined in terms of the potential flow solution (cf. table \ref{tab:geometric_properties}, also \citealp{smeulders_dynamic_1992}).
The first term on the right-hand-side of equation \eqref{eqn:small-time} represents the influence of both the viscous friction and the viscous pressure drag, whilst the subsequent term expresses the net effect of the macroscopic (intrinsic) pressure gradient and the accelerative pressure drag.
The accelerative pressure drag is responsible for redirecting the potential flow solution around the spheres \citep{Graham_2019,Unglehrt2023}, effectively reducing the influence of the intrinsic pressure gradient which drives the flow.
The significance of this pressure drag is accounted by $\alpha_\infty$.
Moreover, one can show that equation \eqref{eqn:small-time} is universal under the viscous normalisation (cf. appendix \ref{app:small-time}).

The current results imply that the general flow features of the start-up flow, consisting of an irrotational (potential) outer flow and thin Stokes boundary layers, is present independent of $\Hg$.
Since the potential outer flow forms immediately after start-up and do not possess any inherent time scale, the flow development during this small-time limit is governed by $\tauen$ (or $\tauvisc$).   
The duration in which the flow follows the solutions of equation \eqref{eqn:small-time} shortens progressively with increasing $\Hg$; In other words, the points where the curves depart from the small-time asymptotics (the dashed lines) shift leftward with increasing $\Hg$.
In relation to the time to reach the steady-state level, however, the high $\Hg$ curves follow the universal asymptotic trajectory actually longer than the lower $\Hg$ curves.
Nonetheless, we have confirmed that the viscous scaling can be used to describe the flow development of nonlinear flow for a small time after the start-up, but the validity breaks down relatively quickly.
As it is discussed next, this observation can be attributed to the emergence of a new time scale taking over the control from $\tauen$. 
 
Thirdly, it is visible in the HCP data that for $\Hg \geq 3.3 \times 10^5$ (Case H3--11) the superficial velocity overshoots before reaching 
the steady-state, whereas for the lower Hagen numbers the superficial velocity relaxes to its steady-state without any overshoot.
In the FCC dataset, the same velocity overshoot can also be observed for $\Hg \geq 5.8 \times 10^4$ (Case F2--11).
Note that this overshoot in a FCC sphere pack was previously mentioned in \citet[][\S 4]{HILL2002a}.
In both sphere-pack configurations, the peaking time normalised with $\tauen$, which characterises the overshoots, decreases significantly with increasing Hagen number (see the peak points shifting leftwards in figures \ref{sub-fig:ubulk-time-hcp-visc} and \ref{sub-fig:ubulk-time-fcc-visc}).
Since the appearance of the velocity overshoot is a unique feature to the nonlinear flow cases, we associate it to the onset of nonlinearity. 
Consequently, it is evident that the viscous time scale does not characterise the onset of this particular nonlinear process.

In the BCC dataset, on the other hand, such temporally localised velocity overshoot cannot be observed.
Instead, for the intermediate to high $\Hg$ cases (B6--11), the levels of $\superficialavg{u}$ that are higher than the corresponding steady-state values persist over significantly longer periods of time before they collapse abruptly (partially visible in figure \ref{sub-fig:ubulk-time-bcc-visc}).

Based on the analysis in \S \ref{sub-sec:onset}, we expect that the inviscid time scale controls the onset of the nonlinearity.
Consequently in figures \ref{sub-fig:ubulk-time-hcp-inv}, \ref{sub-fig:ubulk-time-fcc-inv} and \ref{sub-fig:ubulk-time-bcc-inv}, we present the same data in an inviscid normalisation: $t$ is now normalised by $\tauinv$, and the superficial velocity by $d/\tauinv$ corresponding to equation \eqref{eqn:da-inv}.

In the HCP dataset, the magnitude of the steady-state superficial velocity in the inviscid normalisation seems to converge to a value slightly below $0.1\frac{d}{\tauinv}$ 
for higher Hagen numbers (cf. figure \ref{sub-fig:ubulk-time-hcp-inv}).
In case of FCC, the steady-state amplitudes for the cases with higher $\Hg$ converge around $0.11 \frac{d}{\tauinv}$, whereas the high-$\Hg$ BCC cases match their $\superficialavg{u}$ magnitudes almost perfectly at slightly below $0.25 \frac{d}{\tauinv}$ at $\tauinv=5$, before the aforementioned abrupt drops occur.
Notably, our observation $\superficialavg{u} \propto \frac{d}{\tauinv}$ implies that the drag force becomes asymptotically proportional to the square of the superficial velocity in the high Hagen number limit, which is in accordance with the quadratic Forchheimer correction \citep{Forchheimer1901} as follows:
\begin{equation}
    \superficialavg{u} \propto \frac{d}{\tauinv} \quad\Rightarrow\quad \superficialavg{u}^2 \propto \frac{d^2}{\tauinv^2} = \frac{d \left\vert\bnabla\!\intrinsicavg{p}\right\vert}{\rho} \ .
    \label{eqn:superficial_velocity_scaling}
\end{equation}
This observed scaling is an indication that the macroscopic velocity development around $t \sim \tauinv$ can be considered as an inviscid process in the high $\Hg$ limit. 

\begin{figure}
  \centering
  \subcaptionbox{\label{sub-fig:tpeak_tinv-hcp_fcc}}{
  \includegraphics[width=0.47\textwidth]{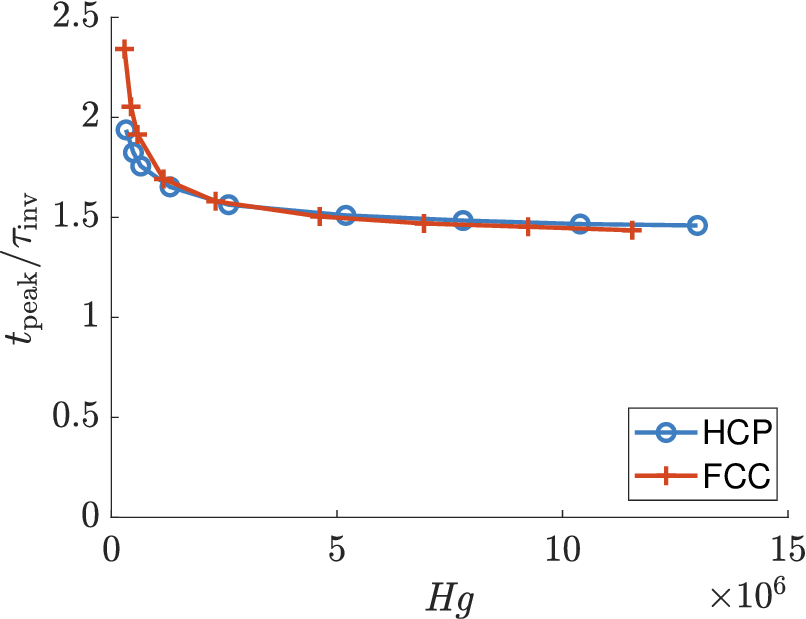}
  } 
  \subcaptionbox{\label{sub-fig:t1pct_hg-hcp_fcc_bcc}}{
  \includegraphics[width=0.48\textwidth]{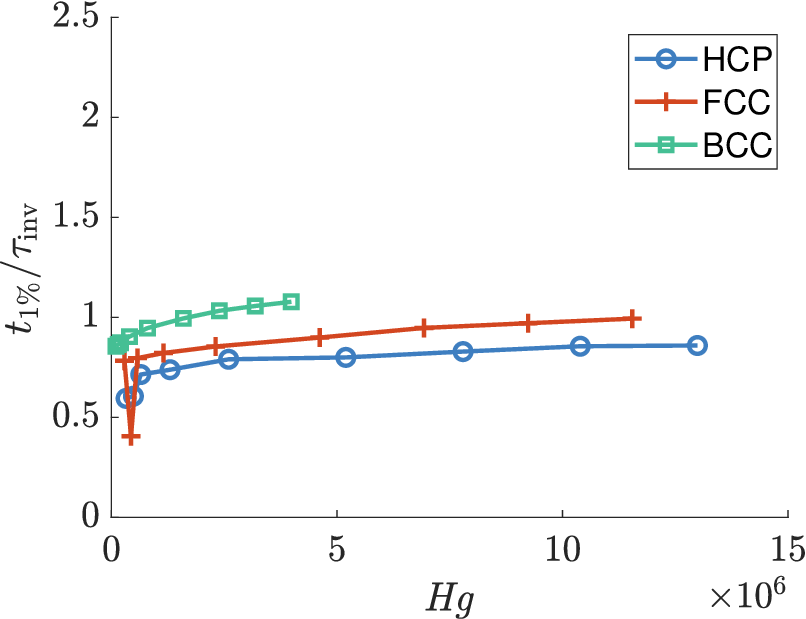} 
  }
  \caption{
  (a) Peaking time ($t_\mathrm{peak}$), and (b) time of departure from the linear flow ($t_\mathrm{1\%}$) normalised by $\tauinv$ vs. $\Hg$
  }
  \label{fig:tpeak_tinv}
\end{figure}

Next, we focus on the scaling of the emergence time of nonlinear events, such as the peaking time $t_{\mathrm{peak}}$.
In the HCP dataset, $t_{\mathrm{peak}}$ converges to a value between $1.4\,\tauinv$ and $1.5\,\tauinv$ (cf. also figures \ref{sub-fig:tpeak_tinv-hcp_fcc}).
The same trend is true for the FCC dataset (cf. figures \ref{sub-fig:ubulk-time-fcc-inv} and \ref{sub-fig:tpeak_tinv-hcp_fcc}), whereas $t_{\mathrm{peak}}$ does not exist in case of the BCC sphere pack as mentioned.
This motivates us to define a more robust measure of emerging nonlinearity.
Correspondingly, figure \ref{sub-fig:t1pct_hg-hcp_fcc_bcc} quantifies the time at which the relative deviation of $\superficialavg{u}$ from the corresponding linear flow solution exceeds $1\,\%$.
Since the linear flow developments from our sphere packs generally follow the small-time asymptotics \eqref{eqn:small-time} up to $t/\tauen \sim 0.5$, whilst the superficial velocities of high $\Hg$ cases reach to their peaks well before this dimensionless time (cf. figures \ref{sub-fig:ubulk-time-hcp-visc}, \ref{sub-fig:ubulk-time-fcc-visc} and \ref{sub-fig:ubulk-time-bcc-visc}), the above departure times also correspond to the departures from the small-time asymptotics. 
These departure times converge around $0.86\,\tauinv$ for HCP,  $1.0\,\tauinv$ for FCC, and $1.1\,\tauinv$ for BCC, respectively.
Overall, the above scaling behaviours indicate that the inviscid time scale $\tauinv$ marks the end of the validity of the small-time asymptotics controlled by $\tauen$, and transition to the intermediate-time asymptotics characterised by the nonlinearity.

\begin{figure}
  \centering
  \subcaptionbox{\label{sub-fig:reypeak_hg-hcp_fcc}}{
  \includegraphics[width=0.47\textwidth]{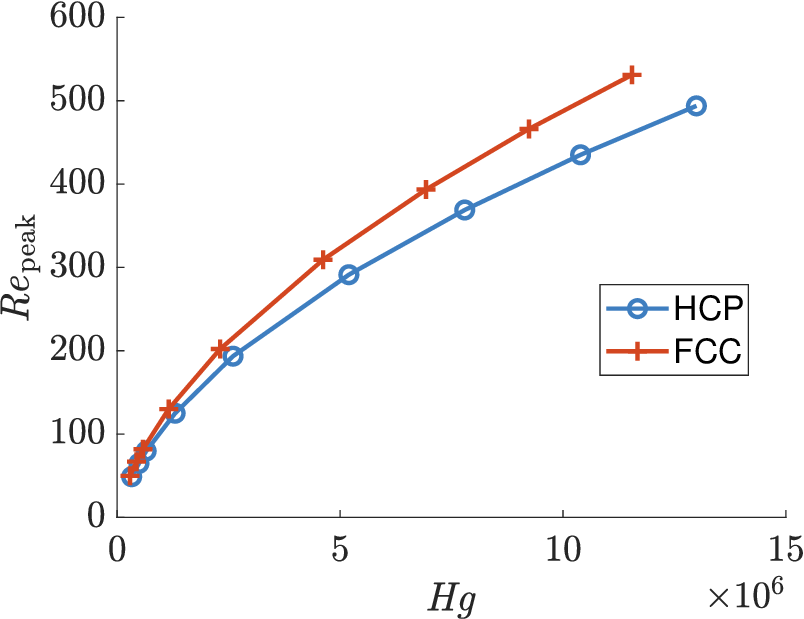}
  } 
  \subcaptionbox{\label{sub-fig:rey1pct_hg-hcp_fcc_bcc}}{
  \includegraphics[width=0.48\textwidth]{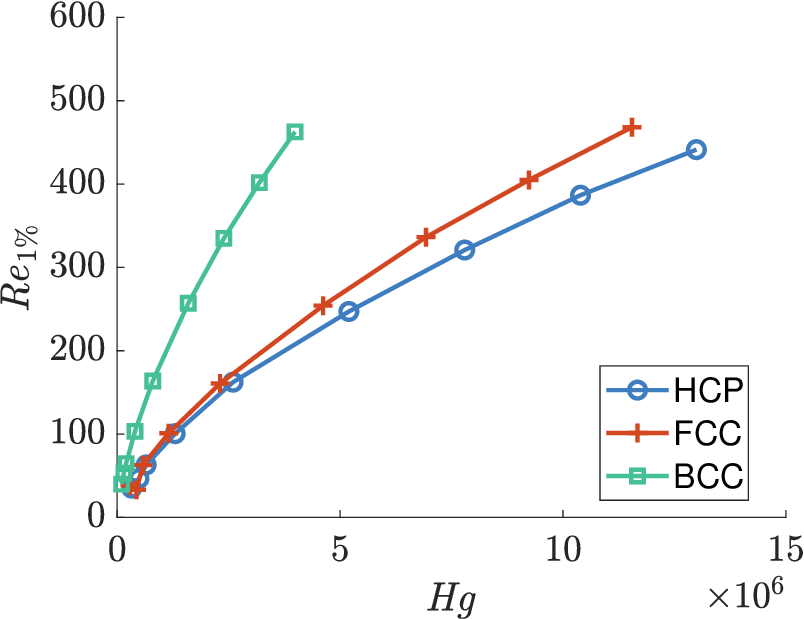} 
  }
  \caption{
  (a) Instantaneous peak Reynolds number ($\Rey_\mathrm{peak}$), and (b) Reynolds number at the time of departure from the linear flow ($\Rey_{\mathrm{1\%}}$) vs. $\Hg$
  }
  \label{fig:reypeak}
\end{figure}

In steady flows, the Reynolds number uniquely identifies the flow regime and can be used to characterise whether flow is linear or nonlinear (cf. \S \ref{sec:intro}). 
Therefore, a question may arise, such as whether it exists a threshold for the instantaneous Reynolds number: 
$\Rey(t)=\superficialavg{u}\!(t)\,d/\nu$ that describes the onset of nonlinear effects in {\em transient} flows. 
Figures \ref{sub-fig:reypeak_hg-hcp_fcc} and \ref{sub-fig:rey1pct_hg-hcp_fcc_bcc} provide the answer to this question by depicting the instantaneous Reynolds number at the peak superficial velocity and at the time when the deviation from the linear evolution exceeds $1\,\%$ as a function of $\Hg$, respectively. 
It is evident that the Reynolds number at which the flow departs from linear behaviour increases with $\Hg$ without reaching an asymptote. 
Therefore, it is demonstrated that there does not exist a fixed value of the instantaneous Reynolds number that marks the departure from the linear flow.

To conclude, the DNS data support that the onset of nonlinearity is indeed timed by the inviscid time scale $\tauinv$ across all three sphere packs being considered.
In other words, in the high-$\Hg$ start-up flows, the nonlinear effects appear at a certain time, not at a certain velocity associated to any specific critical Reynolds number.
This is because, in the high-$\Hg$ limit the inertial effect within the Stokes boundary layers upon the sphere surfaces grows faster than the boundary layer itself (cf. \S \ref{sub-sec:onset}).
In contrast, if $\Hg$ is sufficiently low, the boundary layers grow rapidly and fill entire pore space, not leaving the nonlinearity sufficient time to manifest, resulting that quasi-steady-state assumption to be valid.
   
\section{Pore-scale similarity in nonlinear start-up flow}
\label{sec:similarity}

So far, we have confirmed that the inviscid time scale $\tauinv$ characterises the onset of nonlinear effects in the superficial (macroscopic) velocity at large Hagen numbers.
Based on an order-of-magnitude assessment, we hypothesised that this behaviour is a manifestation of the emerging inertial influence within the boundary layers on the sphere surfaces, which becomes significant at a certain time $t \sim \tauinv$.
In this section, the above hypothesis is confirmed by analysing the pore-scale flow fields.

\subsection{Spatio-temporal development of boundary layer and manifestation of nonlinearity}
\label{sub-sec:streamwise_vorticity_dist}

The acceleration phase of the start-up flow through a sphere pack is characterised by Stokes boundary layers growing under irrotational core flows. 
This implies that we can analyse boundary layer growth by tracking the spatio-temporal development of intense vorticity layers. 
In this analysis, therefore, we start by examining the development of instantaneous streamwise vorticity distributions across a certain cross-flow plane tailored to each sphere pack.
Whilst examining cross-flow sections of sphere-pack porous media flow is a common approach \citep[e.g.][]{HILL2002a,Suekane2003,He2019,Sakai2020}, in this study such analysis is done under a specific assumption that if the flow is similar on the cross-section, it is likely to be similar in the volume as well.
Subsequently, this assumption is confirmed by examining a symmetry plane perpendicular to the cross-flow plane in the HCP sphere pack.  

\begin{figure}
    \centering
    \subcaptionbox{HCP \label{sub-fig:hcp_entry}}{
    \begin{tikzpicture}
        \node [above right,inner sep=0] (image) at (0,0) {
        \includegraphics[width=0.32\linewidth]{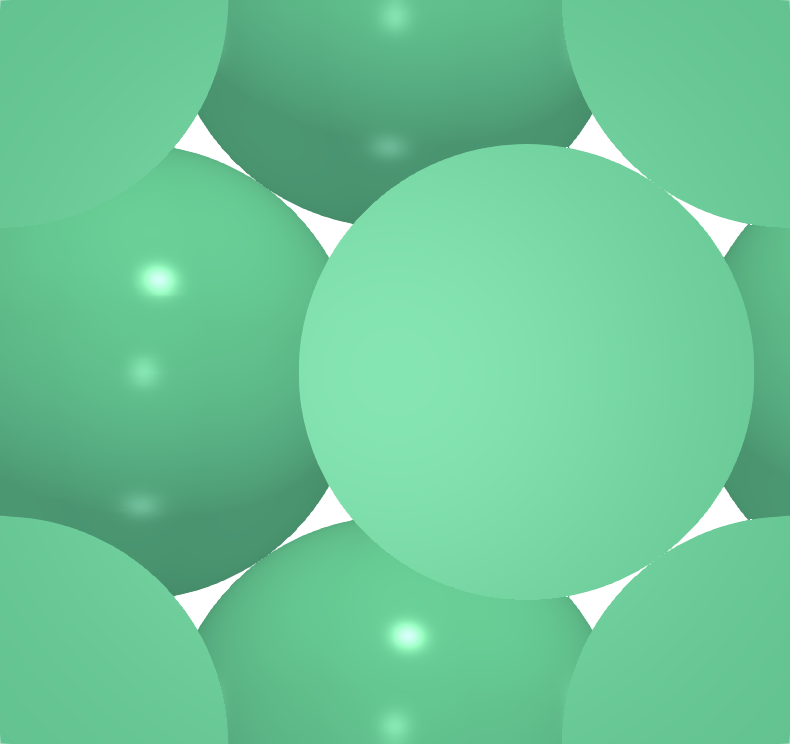}
        };

        \begin{scope}[shift={(image.south west)},
                      x={($(image.south east)-(image.south west)$)},
                      y={($(image.north west)-(image.south west)$)}
                     ]
        \draw[-latex,very thick,black] (0.0,0.0) -- ++(+0.2,0.0) node[pos=1,above]{$y$};
        \draw[-latex,very thick,black] (0.0,0.0) -- ++(+0.0,0.2) node[pos=1,right]{$z$};
        \end{scope}
      \end{tikzpicture}
    }
    \subcaptionbox{FCC \label{sub-fig:fcc_entry}}{
       \includegraphics[width=.31\linewidth]{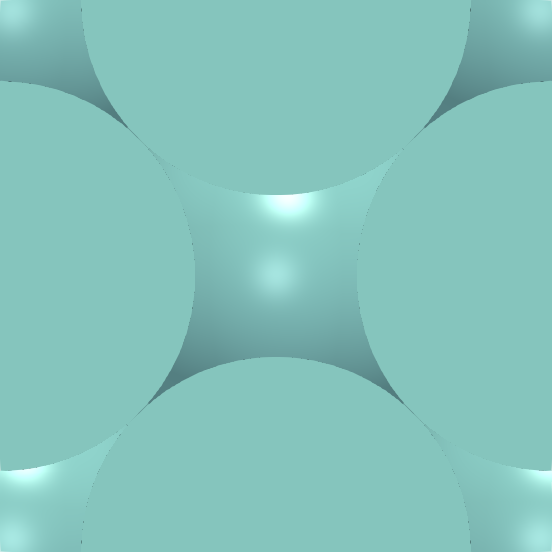}
    }
    \subcaptionbox{BCC\label{sub-fig:bcc_entry}}{
       \includegraphics[width=.31\linewidth]{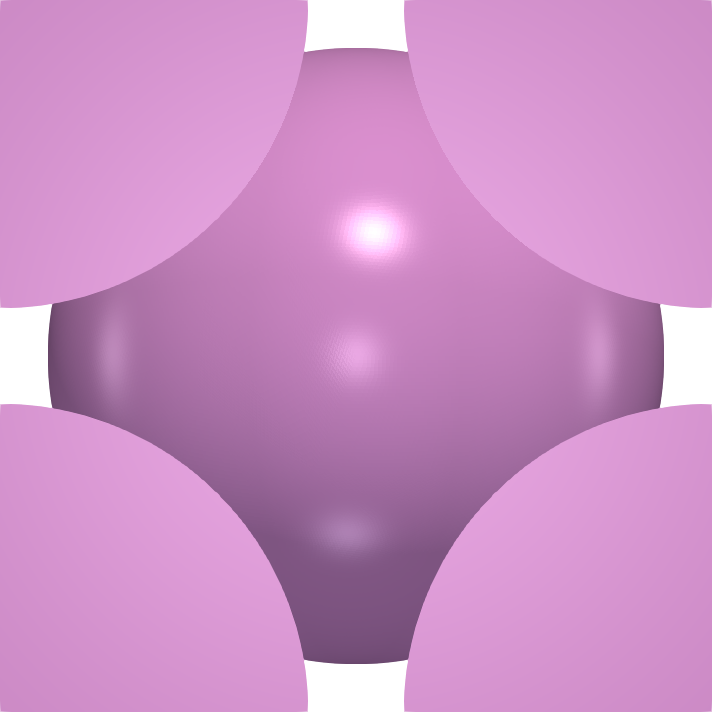}
    }
    \caption{
    Pore structure visualisations from the cross-stream planes shown in figure \ref{fig:geometry}.
    The primary flow direction is towards the reader
    }
    \label{fig:geometry_entryplane}
\end{figure}

For each sphere pack, a sampling plane corresponding to the green shaded rectangles in figure \ref{fig:geometry} (cf. also figure \ref{fig:geometry_entryplane}) was placed. 
The positions of those planes were chosen such that if the pore-scale velocity fields maintain the fore-aft symmetry, the cross-plane velocity as well as the streamwise vorticity components on the planes are zero. 
Since this fore-aft symmetry is a signature of the linear flow processes \citep{Unglehrt2022}, the development of non-zero $\omega_x$ on these sampling planes can be seen as a direct indicator of the emerging nonlinearity.
From this viewpoint, the consistent development of pore-scale vortical flow structures in transient nonlinear flow, observed in \citet{Sakai2020} for the HCP geometry, is a nonlinear phenomenon itself.

\begin{figure}
    \centering
    \subcaptionbox{$t/\tauinv=0.6$\label{fig:streamwise_vorticity_tauinv06_HCP}}{
    \includegraphics[width=\textwidth]{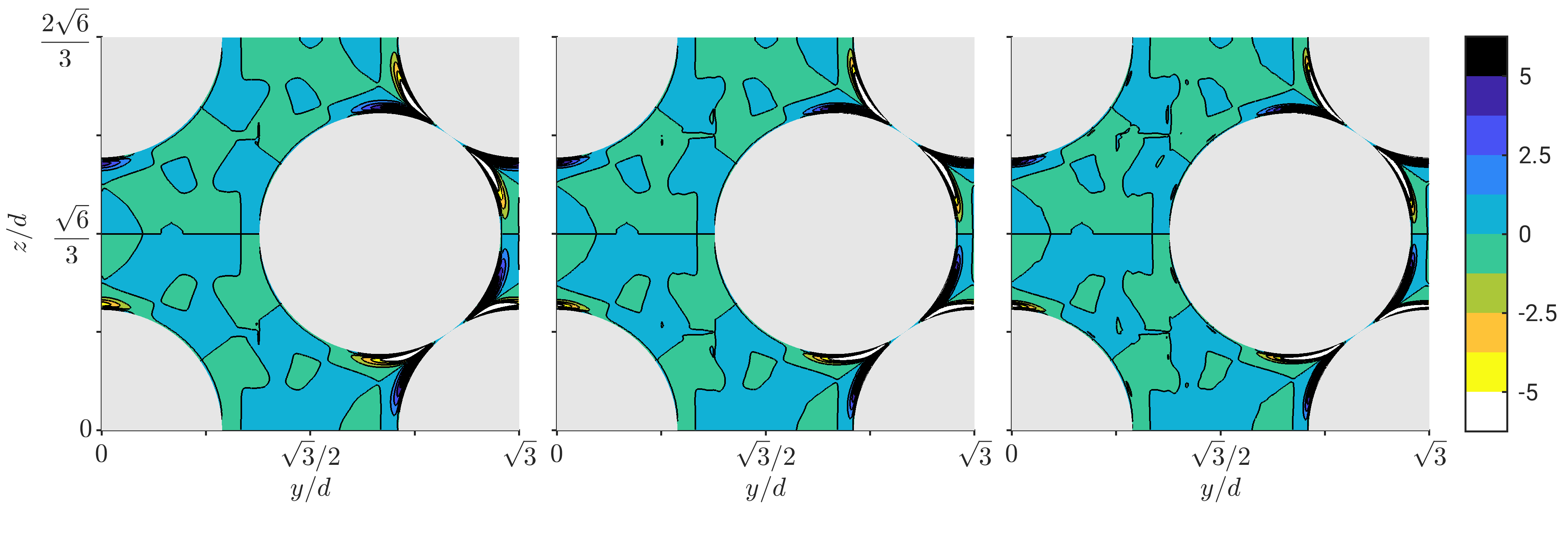}
    }
    \subcaptionbox{$t/\tauinv=1.0$\label{fig:streamwise_vorticity_tauinv10_HCP}}{
    \includegraphics[width=\textwidth]{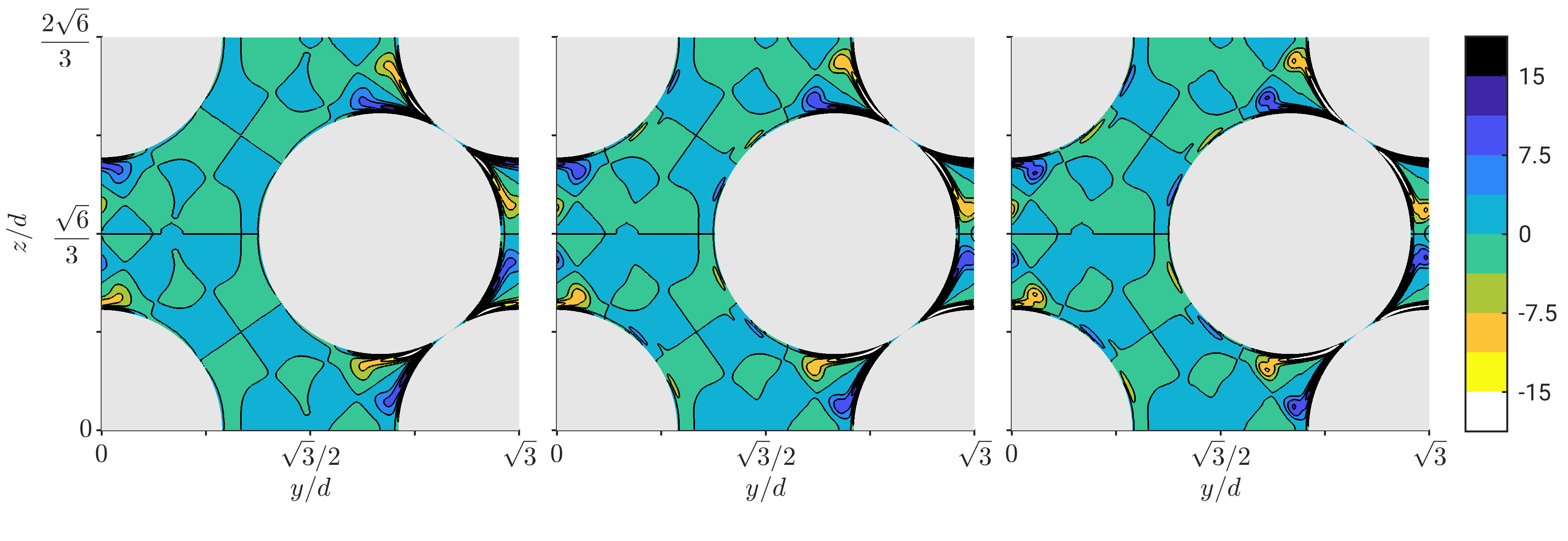}
    }
    \subcaptionbox{$t/\tauinv=1.4$\label{fig:streamwise_vorticity_tauinv14_HCP}}{
    \includegraphics[width=\textwidth]{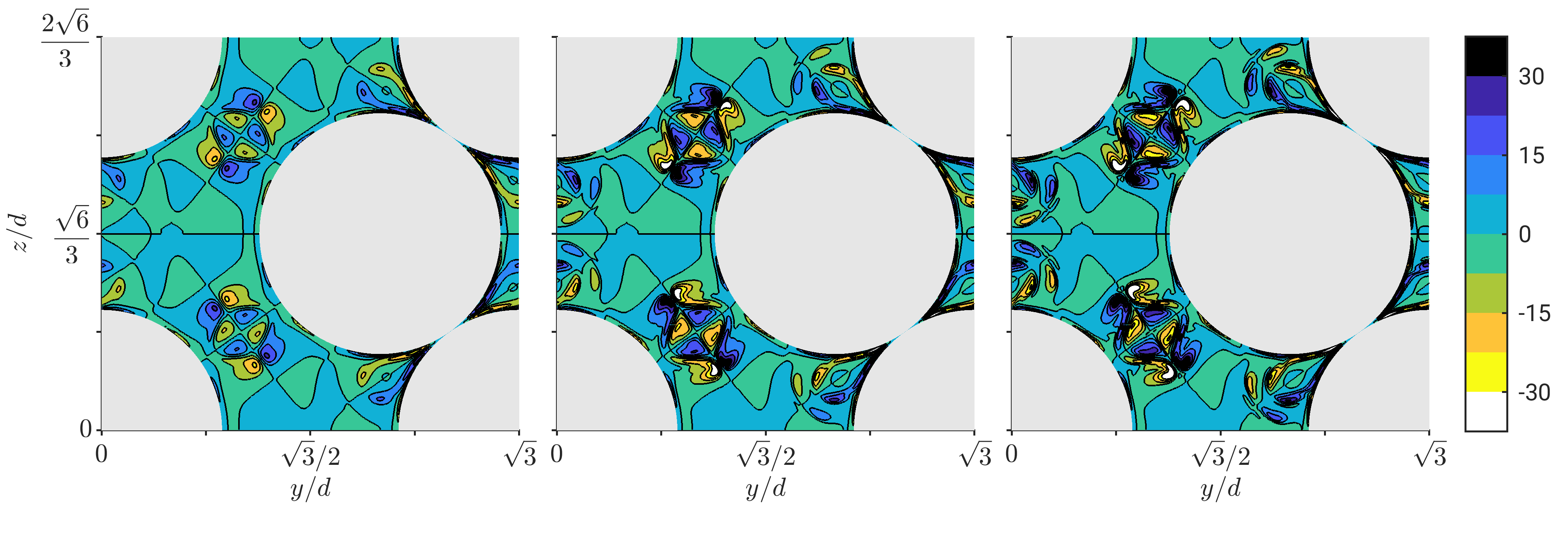}
    }
    \caption{
    Instantaneous streamwise vorticity $\omega_x \tauinv$ on the sampling plane of HCP. 
    left column: H7 at $\Hg=2.6\times 10^{6}$; 
    middle column: H9 at $\Hg = 7.8\times 10^{6}$; 
    right column: H11 at $\Hg = 1.3 \times 10^7$
    }
    \label{fig:streamwise_vorticity_tauinv_HCP}
\end{figure}

\begin{figure}
    \centering
    \subcaptionbox{$t/\tauinv=0.6$\label{fig:streamwise_vorticity_tauinv06_FCC}}{
    \includegraphics[width=\textwidth]{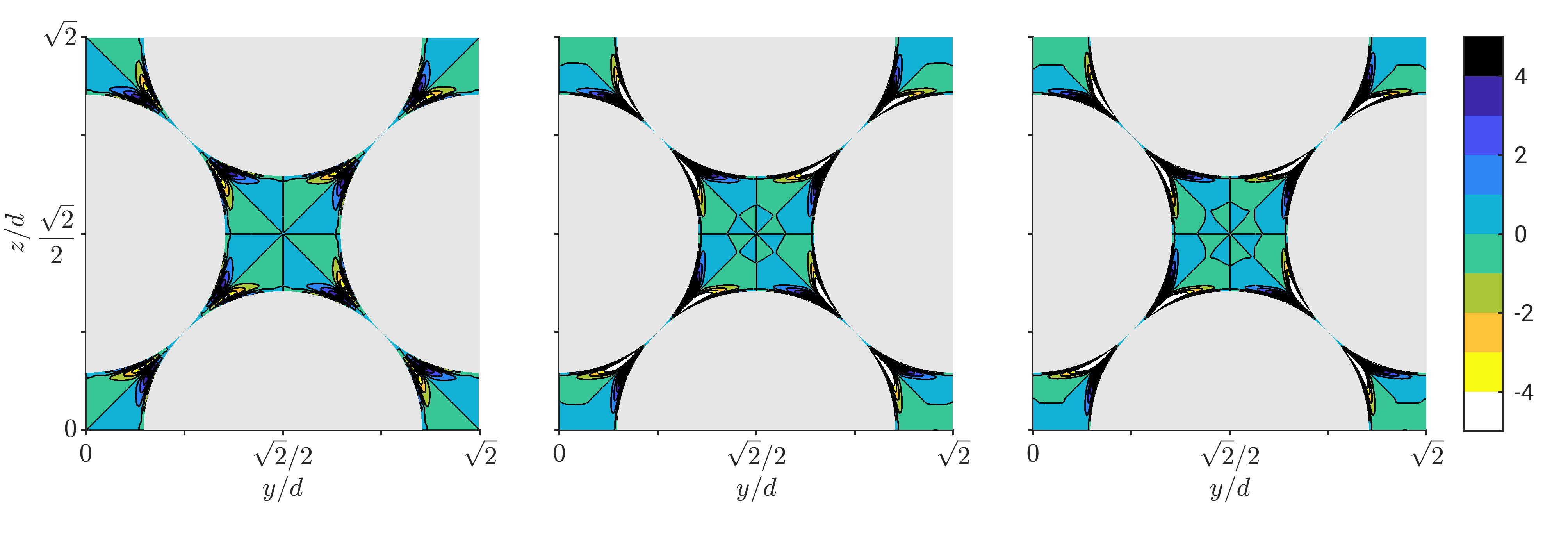}
    }
    \subcaptionbox{$t/\tauinv=1.0$\label{fig:streamwise_vorticity_tauinv10_FCC}}{
    \includegraphics[width=\textwidth]{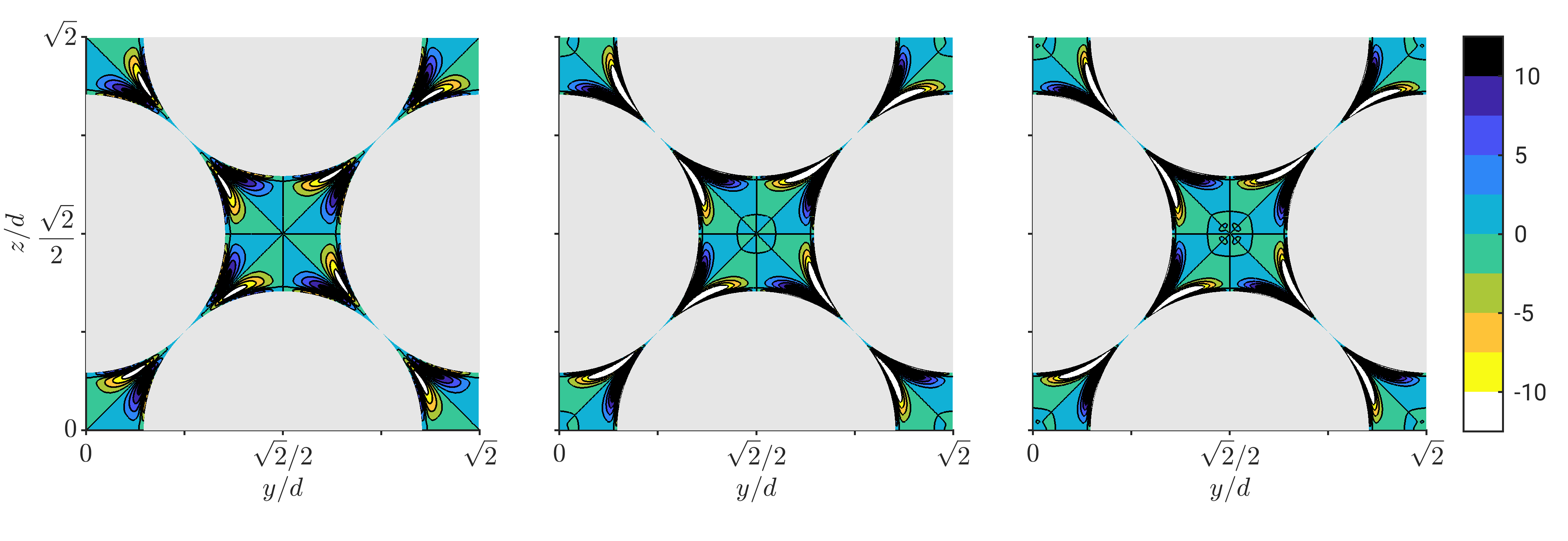}
    }
    \subcaptionbox{$t/\tauinv=1.4$\label{fig:streamwise_vorticity_tauinv14_FCC}}{
    \includegraphics[width=\textwidth]{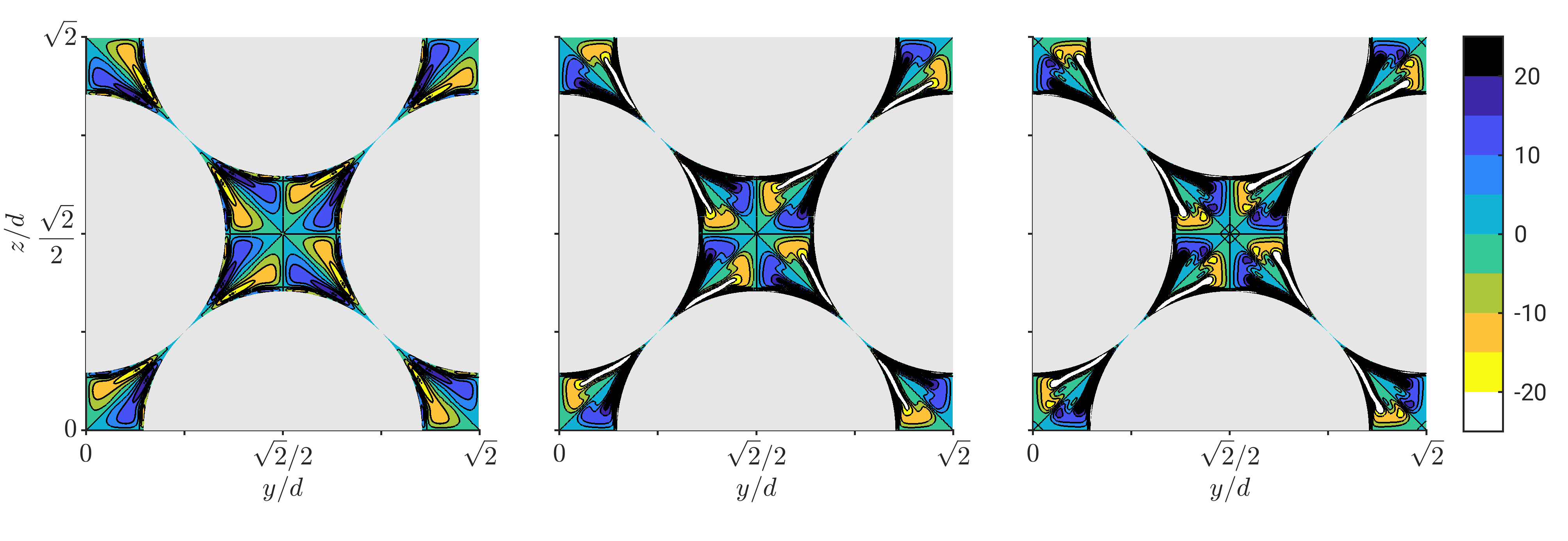}
    }
    \caption{
    Instantaneous streamwise vorticity $\omega_x \tauinv$ on the sampling plane of FCC. 
    left column: F7 at $\Hg = 2.3 \times 10^6$; 
    middle column: F9 at $\Hg = 6.9 \times 10^6$; 
    right column: F11 at $\Hg = 1.2 \times 10^7$.
    }
    \label{fig:streamwise_vorticity_tauinv_FCC}
\end{figure}

\begin{figure}
    \centering
    \subcaptionbox{$t/\tauinv=0.6$\label{fig:streamwise_vorticity_tauinv06_BCC}}{
    \includegraphics[width=\textwidth]{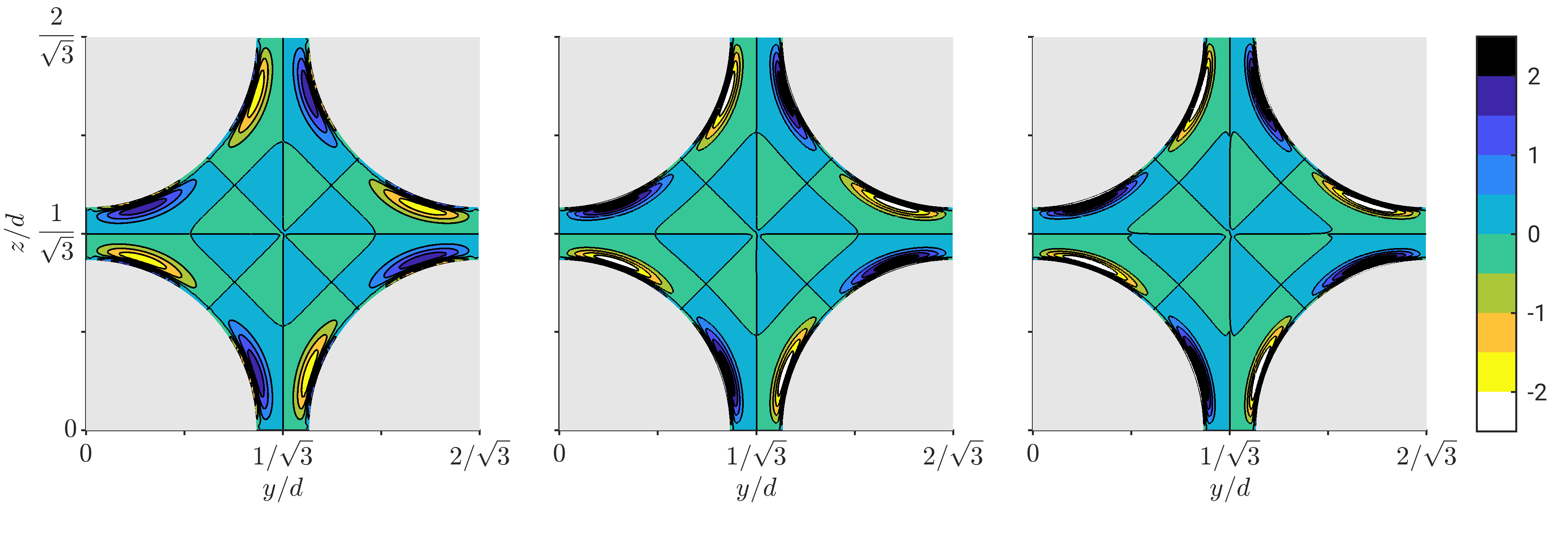}
    }
    \subcaptionbox{$t/\tauinv=1.0$\label{fig:streamwise_vorticity_tauinv10_BCC}}{
    \includegraphics[width=\textwidth]{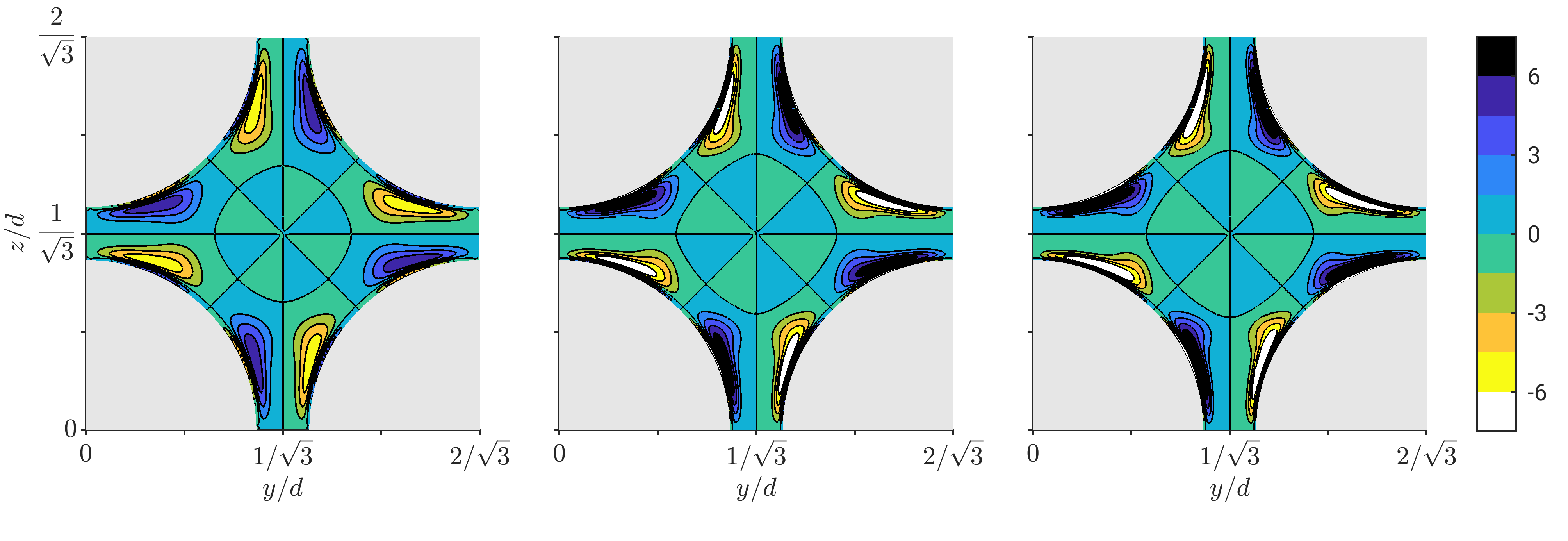}
    }
    \subcaptionbox{$t/\tauinv=1.4$\label{fig:streamwise_vorticity_tauinv14_BCC}}{
    \includegraphics[width=\textwidth]{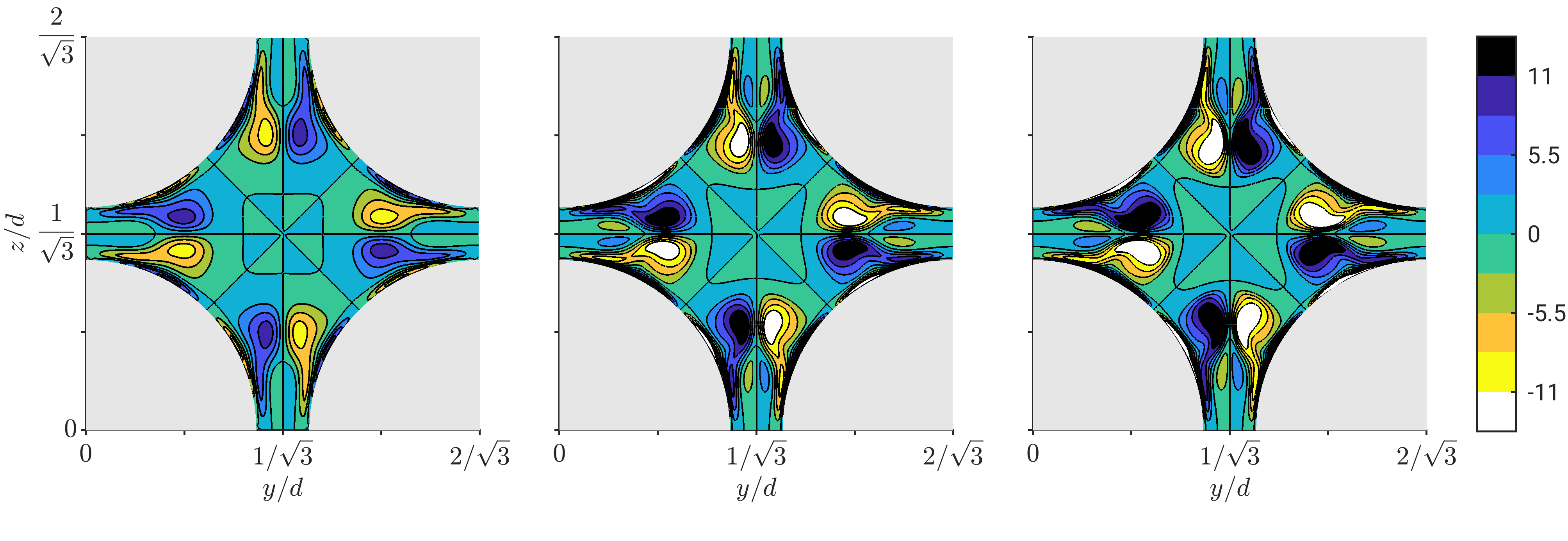}
    }
    \caption{
    Instantaneous streamwise vorticity $\omega_x \tauinv$ on the sampling plane of BCC. 
    left column: B7 at $\Hg = 8 \times 10^5$; 
    middle column: B9 at $\Hg = 2.4 \times 10^6$; 
    right column: B11 at $\Hg = 4 \times 10^6$
    }
    \label{fig:streamwise_vorticity_tauinv_BCC}
\end{figure}

Figures \ref{fig:streamwise_vorticity_tauinv_HCP}, \ref{fig:streamwise_vorticity_tauinv_FCC}, \ref{fig:streamwise_vorticity_tauinv_BCC} depict $\omega_x \tauinv$ distributions on the respective sampling planes at three time instances $t/\tauinv=\{0.6,\,1.0,\,1.4\}$. 
These dimensionless instances with respect to $\tauinv$ lie within the development phase of the nonlinear effects based on $t_\mathrm{peak}/\tauinv$ and $t_{1\,\%}/\tauinv$ (cf. figure \ref{fig:tpeak_tinv}).
Three cases of each sphere-pack geometry were chosen, namely: H7, 9, 11 (HCP); F7, 9, 11 (FCC); and B7, 9, 11 (BCC), respectively.
Within each set of cases, the highest $\Hg$ is five times of the lowest $\Hg$, whilst the median $\Hg$ is 60\% of the highest $\Hg$ (cf. tables \ref{tab:hcp-param} and \ref{tab:fcc-bcc-param}).
Here, the highest and the second highest $\Hg$ cases for each sphere pack represent the high-$\Hg$ limit behaviours, whereas the lowest $\Hg$ cases showcase possible low-$\Hg$ effects. 

Generally among the three sphere packs, the normalised instantaneous vorticity contours are virtually indistinguishable both qualitatively and quantitatively between the two high-$\Hg$ limit cases.
Only exceptions to the observed similarity are the magnitude of the peaks residing inside the highly localised areas, which grows slowly with increasing $\Hg$, as well as the thickness of the boundary layers that becomes slightly thinner with $\Hg$.
Even at the lowest Hagen numbers, the pore-scale flow developments are qualitatively similar, whilst clear signs of low-$\Hg$ effects are more noticeable in the peak magnitude of the vorticity and the boundary layer thickness. 

At the earliest time of $t/\tauinv=0.6$, it can be clearly seen that strong vorticity layers exist directly above the sphere surfaces. 
In the cases of HCP and FCC, those layers are mainly located around the sphere contact points on the respective sampling plane, whilst such peaks reside in the proximity of the four unobstructed passages in the cases of BCC (cf. figure \ref{fig:geometry_entryplane}).
The vorticity magnitude in the rest of the pores are essentially zero, confirming the existence of the ideal potential core flows at this dimensionless time.
Subsequently, those vorticity boundary layers grow in their thickness via viscous diffusion, which is particularly visible at $t/\tauinv=1.0$ in FCC and BCC (cf. figures \ref{fig:streamwise_vorticity_tauinv10_FCC} and \ref{fig:streamwise_vorticity_tauinv10_BCC}).

At the latest time ($t/\tauinv=1.4$), the strong vorticity layers detach from the sphere surfaces and migrate into the core flow region, signalling a qualitative shift from the boundary-layer-potential-flow characteristics.
Generally within the high-$\Hg$ flows through HCP, the appearance of the eight vortices in the large octahedral pores (cf. emerging vortex structures in figure \ref{fig:streamwise_vorticity_tauinv14_HCP}) is the consequence of this vorticity boundary layer detachment happening in the preceding tetrahedral pores \citep[cf.][also figure \ref{sub-fig:hcp_entry} for a visualisation of tetrahedral pores through the corresponding octahedral pores]{Sakai2020}.
Our results show that the transport of the detached vorticity layers is controlled by $\tauinv$. 
Finally, it is worth noting that vorticity distributions agree with the geometric symmetries being imposed by the sphere packs in all cases, indicating that the flows remain laminar even at this later time.

\begin{figure}
    \centering
    \subcaptionbox{$t/\tauinv=0.6$\label{fig:velocity-cutplane_tauinv06_HCP}}{
    \begin{tikzpicture}
    
    \node [above right,inner sep=0] (image) at (0,0) {
        \includegraphics[width=.9\textwidth]{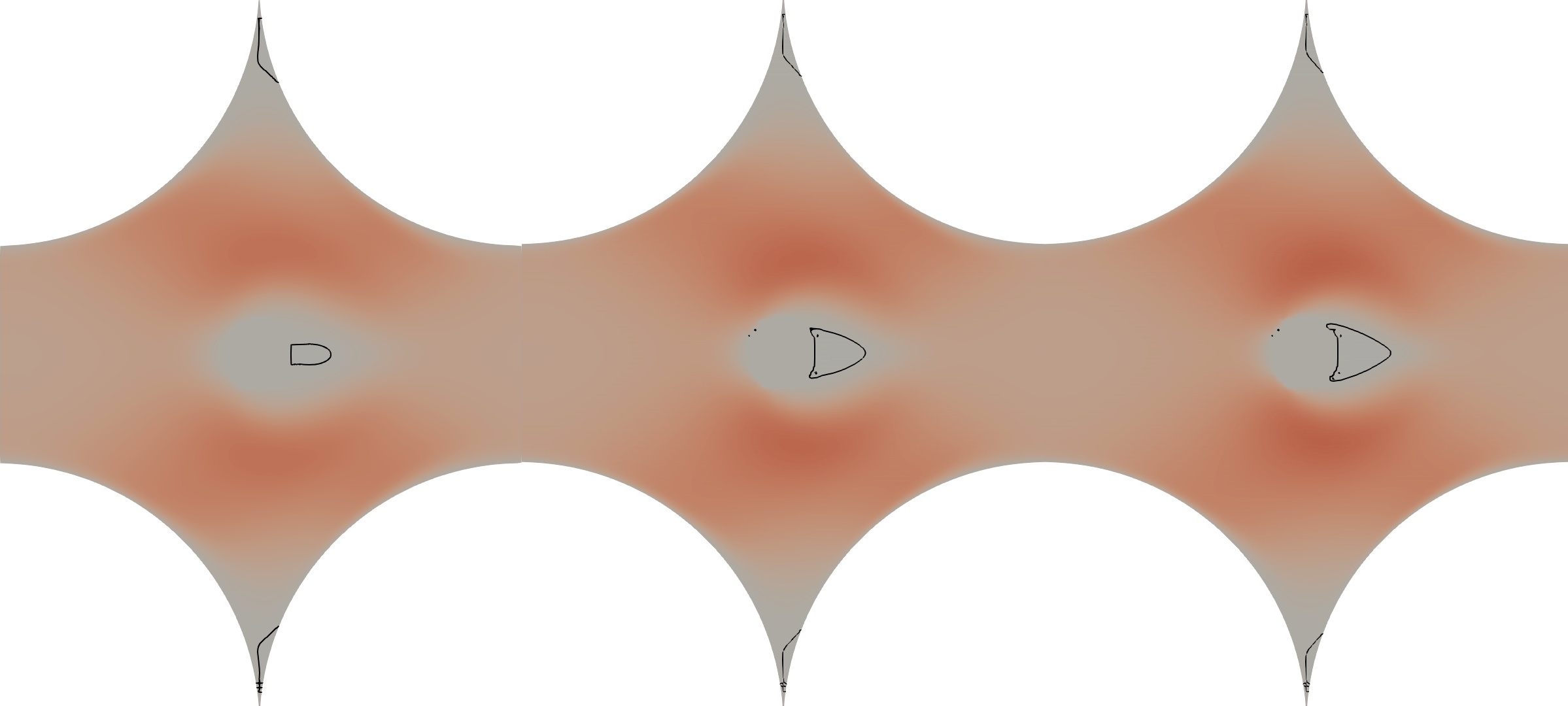}};
    \begin{scope}[
    x={($.032*(image.south east)$)},
    y={($.1*(image.north west)$)}]
 
       \draw[latex-,very thick,black] (5,5) -- ++(-3,-3) node[below,black]{\small contact point};
    \end{scope}
    \end{tikzpicture}
    }
    \subcaptionbox{$t/\tauinv=1.0$\label{fig:velocity-cutplane_tauinv10_HCP}}{
    \includegraphics[width=.9\textwidth]{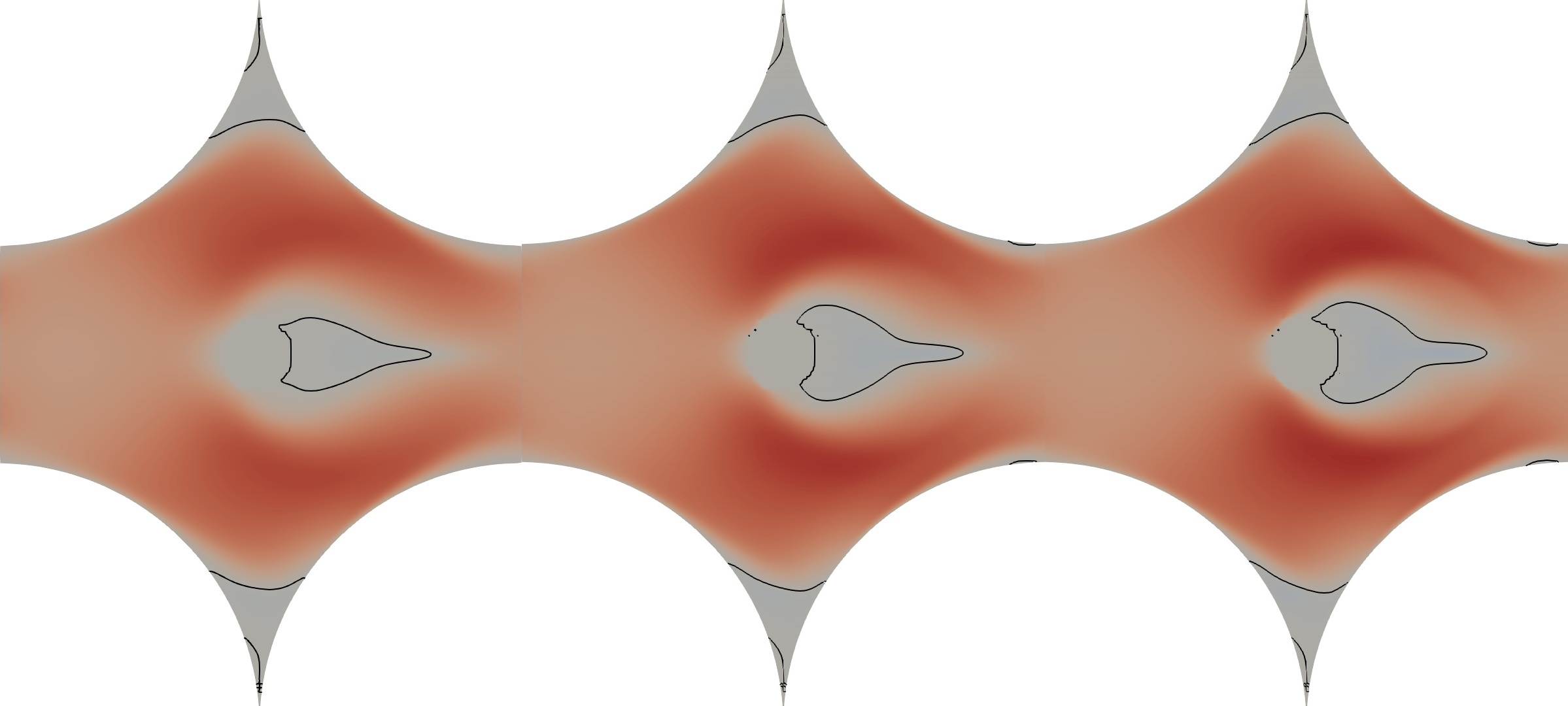}
    }
    \subcaptionbox{$t/\tauinv=1.4$\label{fig:velocity-cutplane_tauinv14_HCP}}{
    \includegraphics[width=.9\textwidth]{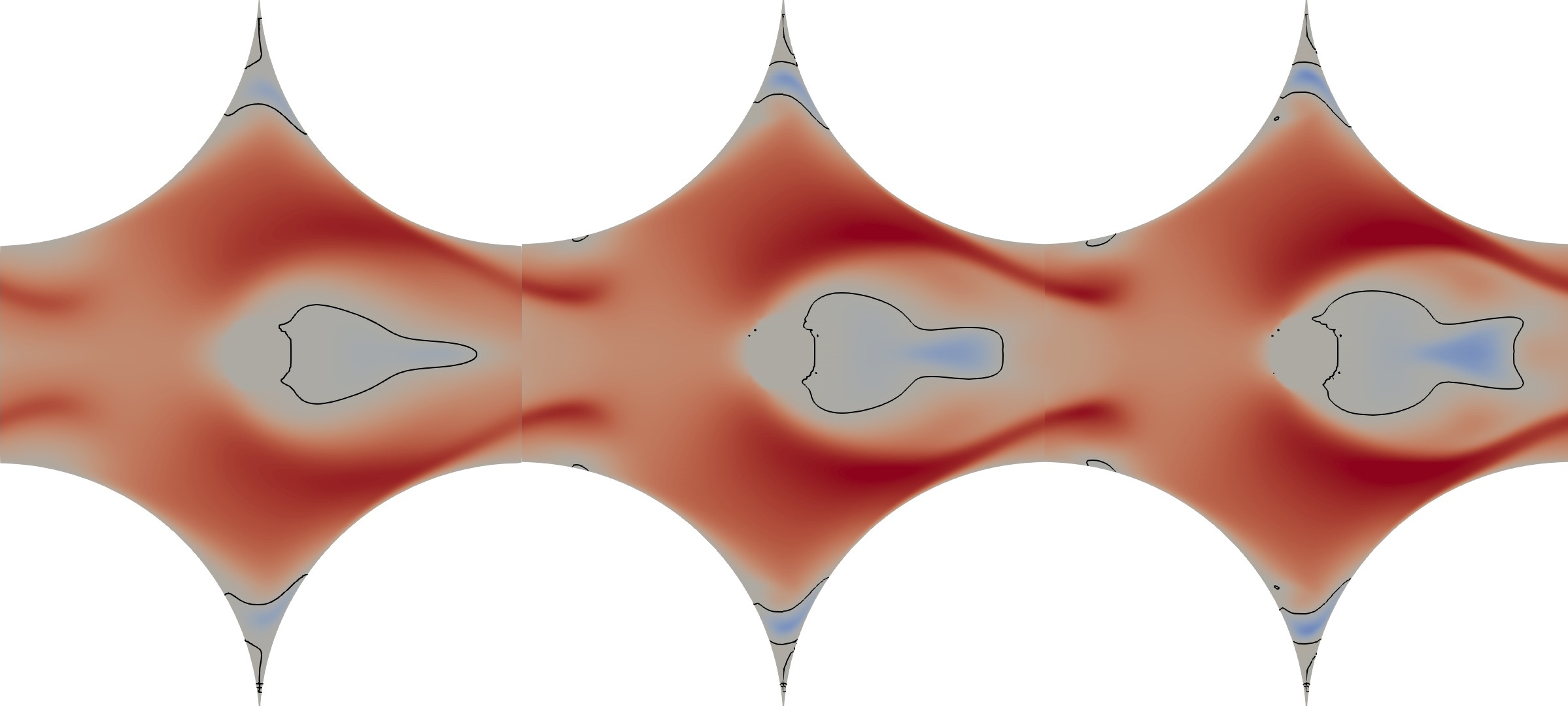}
    }
    \includegraphics[width=0.25\linewidth]{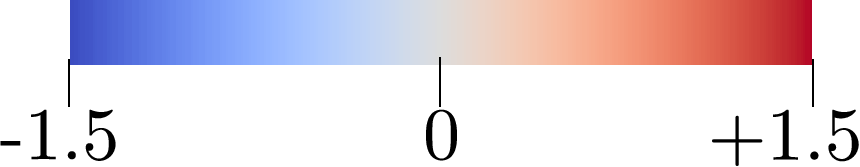}
    \caption{
    Instantaneous streamwise velocity $u\, \tauinv/d$ 
    (colour) being annotated by zero-velocity contour lines (black) on a local symmetry plane in the HCP sphere pack.
    The primary flow direction is from left to right, and only one half of the computational domain in the $x$-direction per column is shown for better visibility.
    left column: H7; middle column: H9; right column: H11
    }
    \label{fig:velocity-cutplane}
\end{figure}

Above vorticity layer detachment is directly linked to flow separations from the sphere surfaces, and is the underlying formation mechanism of so-called \textit{inertial core} that have been reported repeatedly over the past \citep{Dybbs1984,HILL2001a,Horton2009,Sakai2020,Forslund2021}.
To examine this separation process, the corresponding instantaneous velocity fields on a local symmetry plane in the HCP sphere pack, defined by $\frac{\sqrt{3}}{3} y -\frac{\sqrt{6}}{3}z=0$, are shown in figure \ref{fig:velocity-cutplane}.
In this figure, the flow enters from left through an octahedral pore with a larger cross section, and is subsequently squeezed in the plane-normal direction in the proximity of the touching spheres (annotated as \textit{contact point}).
The contact points split the flow into two streams, which meet in the following octahedral pore, forming a split-and-merge sequence that repeats in $x$-direction.
When $\Hg$ is sufficiently high, the two high-momentum streams cannot remain attached to the sphere surfaces, forming the separation bubbles visible in the downstream of the contact points, even at $t/\tauinv=0.6$.
Subsequently, the separation bubbles grow in size with increasing time, whilst the two streams laterally separate from one another, eventually resulting a total detachment of the high-momentum jets from the sphere surfaces (i.e. inertial core formation) at $t/\tauinv=1.4$. 
During this process, the velocity fields remain qualitatively and quantitatively similar between the two high-$\Hg$ cases, especially at $t/\tauinv = \{0.6, 1.0\}$, whilst the low-$\Hg$ also exhibits a qualitative similarity.
At the latest time at $t/\tauinv = 1.4$, minor qualitative $\Hg$-dependence becomes noticeable on the tale end of the separation bubbles.
In short, we observe significant flow separations inside the HCP sphere pack, and they coincide with the aforementioned vorticity layer detachment.
Most importantly, the emergence of the detachments and the propagation of the separation bubbles are controlled by $\tauinv$, independent of $\Hg$.

To summarise, we have found that consistent evolutions of pore-scale vortical structures also exist in the FCC and the BCC sphere packs ---albeit their patterns are specific to the respective sphere pack geometries--- indicating certain generality of this kind of phenomena which was first found for HCP \citep{Sakai2020}.
Based on the geometric symmetry of the sphere packs, the consistent flow evolution can be seen as an emergence process of the underlying nonlinearity, which initially grows inside the boundary layer on the sphere surfaces.
Across the considered sphere packs, the inviscid time scale $\tauinv$ controls this pore-scale flow evolution, whilst its vorticity magnitude exhibits a weak Hagen number dependency under the inviscid normalisation $\omega_x \tauinv$.
At later times, the growing nonlinearity forces the boundary layer to be detached from the sphere surfaces, spreading nonlinearity into the core flow regions.

\subsection{Laminar and turbulent scaling of streamwise vorticity magnitude}
\label{sub-sec:streamwise_vorticity_magnitude}

Here, we quantify the observed $\Hg$-dependency of the peak $\omega_x$ intensity, and discuss the underlying physical mechanisms based on the identified scaling behaviours.
From the dimensional viewpoint, the vorticity can be treated as a velocity-length ratio, i.e. $\omega_x \sim U/\delta$, where $U$ is the characteristic velocity scale and $\delta$ is the characteristic boundary layer thickness.
As shown in \S \ref{sub-sec:nonlinear}, the velocity development up to around $t \sim \tauinv$ in high-$\Hg$ flow is essentially an inviscid process with approximately a constant acceleration, with the inviscid time scale $\tauinv$ controlling the flow development.
We can, therefore, approximate the velocity scale at $t = \tauinv$ from the potential flow solution as in \S \ref{sub-sec:onset}: $U\sim |\Vec{\nabla}\!\intrinsicavg{p}| \tauinv/\rho$.
Beneath this potential core flow, laminar boundary layers grow around the sphere surfaces with the thickness of $\delta \sim \sqrt{\nu \tauinv}$.   
Combining the two characteristic scales leads to the following vorticity-scaling conjecture:
\begin{equation}
        \omega_x\,\tauinv \sim \frac{|\Vec{\nabla}\!\intrinsicavg{p}|\tauinv}{\rho \sqrt{\nu \tauinv}}\, \tauinv\sim \frac{d}{\sqrt{\nu \tauinv}} \sim \sqrt{\frac{\tau_\mathrm{visc}}{\tauinv}} \sim \Hg^{1/4} \ .
        \label{eqn:omx_laminar_bl_scaling}
\end{equation}

\begin{figure}
    \centering
    \subcaptionbox{HCP in laminar scaling \label{sub-fig:streamwise_vorticity_laminar_scaling_HCP}}{
        \includegraphics[width=0.45\textwidth]{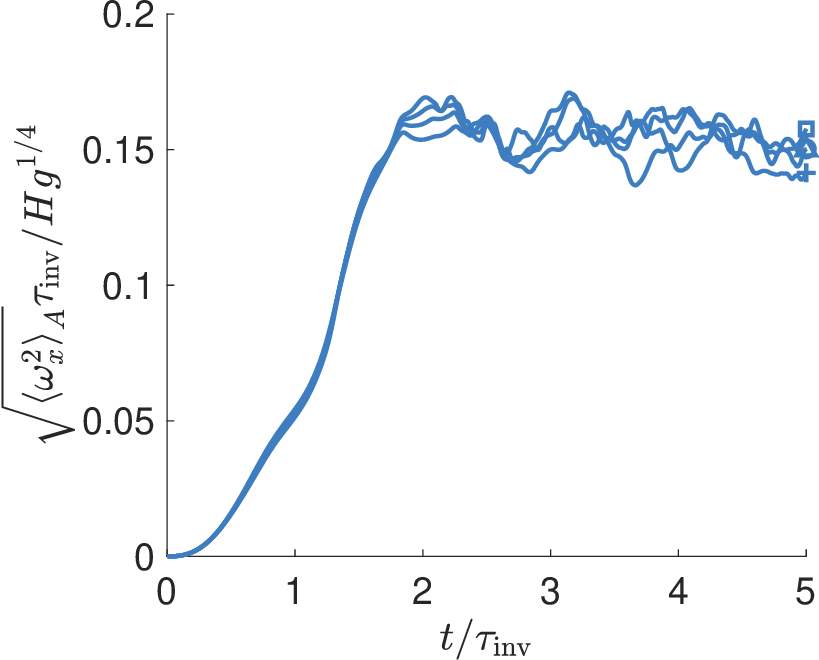}
    }
    \subcaptionbox{HCP in turbulent scaling \label{sub-fig:streamwise_vorticity_turbulent_scaling_HCP}}{
        \includegraphics[width=0.45\textwidth]{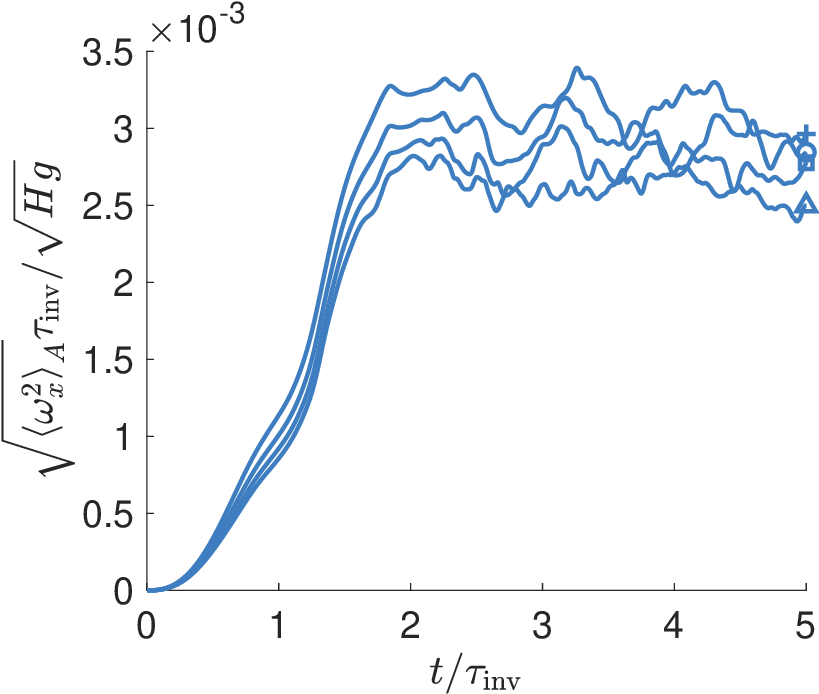}
    }%
    
    \subcaptionbox{FCC in laminar scaling \label{sub-fig:streamwise_vorticity_laminar_scaling_FCC}}{
        \includegraphics[width=0.45\textwidth]{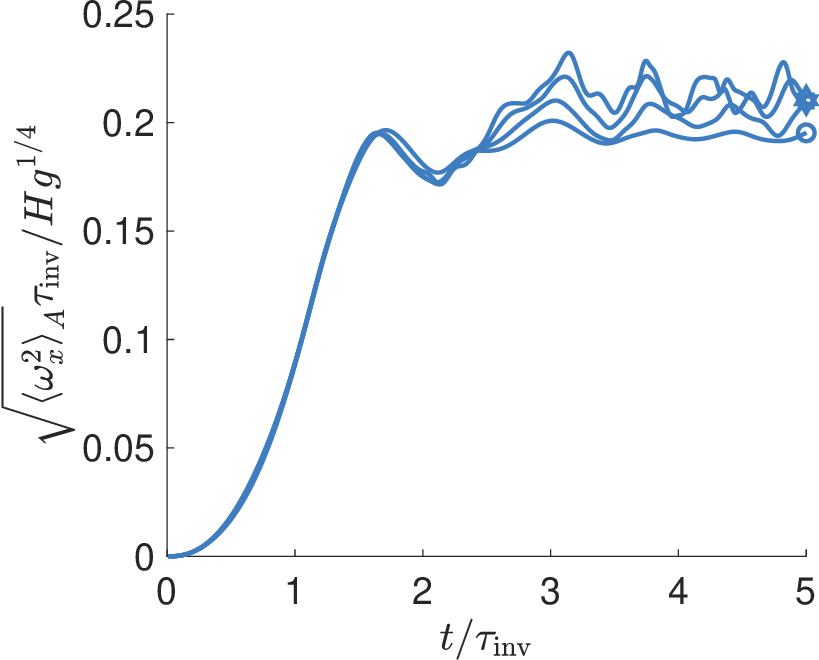}
    }
    \subcaptionbox{FCC in turbulent scaling \label{sub-fig:streamwise_vorticity_turbulent_scaling_FCC}}{
        \includegraphics[width=0.45\textwidth]{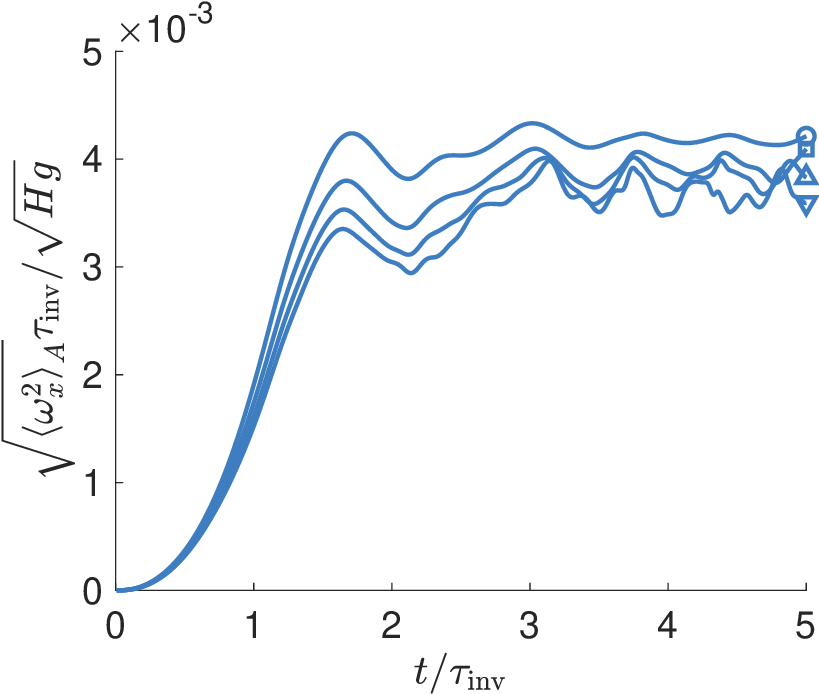}
    }%
    
    \subcaptionbox{BCC in laminar scaling \label{sub-fig:streamwise_vorticity_laminar_scaling_BCC}}{
        \includegraphics[width=0.45\textwidth]{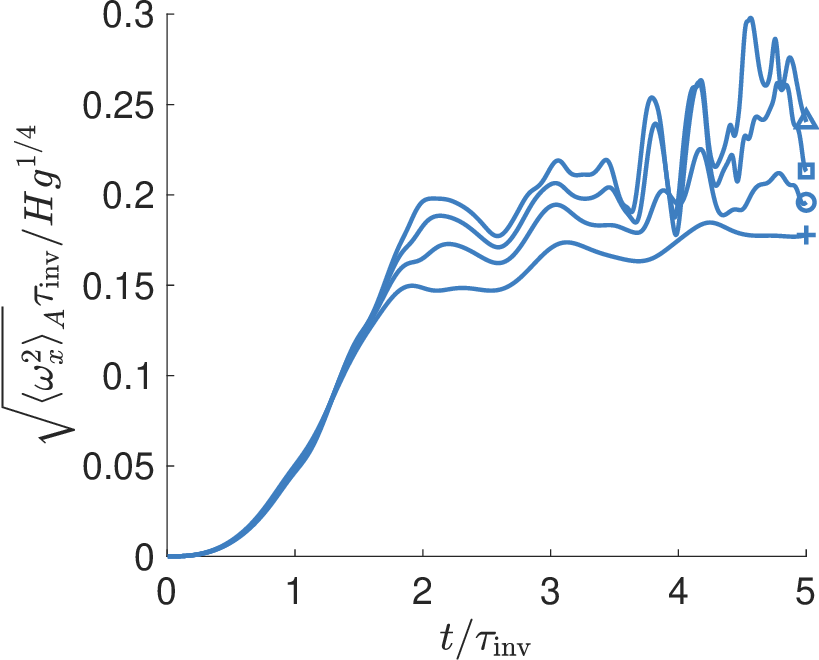}
    }
    \subcaptionbox{BCC in turbulent scaling \label{sub-fig:streamwise_vorticity_turbulent_scaling_BCC}}{
        \includegraphics[width=0.45\textwidth]{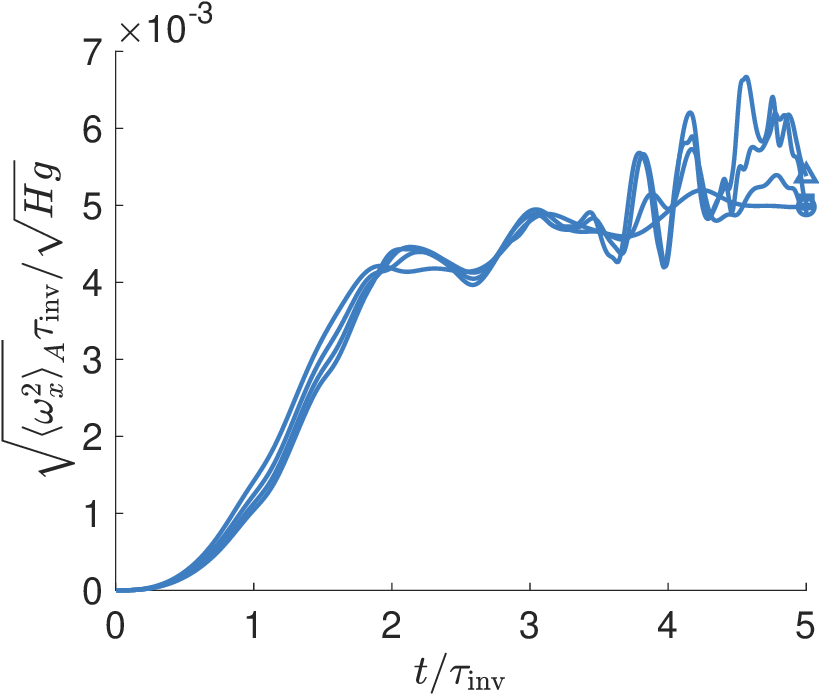}
    }
    \caption{Root-mean-square of plane-averaged streamwise vorticity $\sqrt{\langle\omega_x^2\rangle_{A}}$ in the laminar boundary layer and the turbulent scaling.
    The plotted cases are: H8--11 for HCP, F8--11 for FCC, B8--11 for BCC, respectively 
    (the markers are according to tables \ref{tab:hcp-param} and \ref{tab:fcc-bcc-param}).
    }
    \label{fig:streamwise_vorticity_scaling}
\end{figure}

Consequently, we examine the four highest $\Hg$ cases from each sphere pack (H8--11 for HCP; F8--11 for FCC; B8--11 for BCC), and depict their temporal evolutions of root-mean-square (RMS) of streamwise vorticity on the sampling planes $\sqrt{\langle \omega_x^2\rangle_A}$ in figure \ref{fig:streamwise_vorticity_scaling}.
Note that the streamwise vorticity distributions maintain the geometric symmetries even at $t/\tauinv=1.4$ (cf.  \S \ref{sub-sec:streamwise_vorticity_dist}), therefore the RMS value was selected for quantification in order to circumvent the net-zero values.
In figures \ref{sub-fig:streamwise_vorticity_laminar_scaling_HCP}, \ref{sub-fig:streamwise_vorticity_laminar_scaling_FCC} and \ref{sub-fig:streamwise_vorticity_laminar_scaling_BCC}, the vorticity RMS values are scaled with $\Hg^{1/4}$  according to the laminar boundary layer scaling conjecture \eqref{eqn:omx_laminar_bl_scaling}, whereas a different scaling is applied to figures \ref{sub-fig:streamwise_vorticity_turbulent_scaling_HCP}, \ref{sub-fig:streamwise_vorticity_turbulent_scaling_FCC} and \ref{sub-fig:streamwise_vorticity_turbulent_scaling_BCC}, as detailed in the following discussion.

First, it is clearly visible across all three sphere packs that the streamwise vorticity RMS values lines up, peaks and troughs at same inviscid times, strongly indicating the decisive role of $\tauinv$ in the pore-scale flow development during the flow acceleration phase.
Second, at small (i.e. $t \ll \tauinv$) to intermediate ($t \sim \tauinv$) times, we observe a collapse under the laminar boundary-layer scaling (up to $t/\tauinv\approx 2$ for HCP, $t/\tauinv\approx 2.5$ for FCC, $t/\tauinv \approx 1.8$ for BCC), confirming the validity of this scaling during these phases.
Finally, notice for the BCC sphere pack that the boundary-layer scaling is superseded by so-called \textit{turbulent} scaling ---dividing the vorticity RMS by $\Hg^{1/2}$ instead of $Hg^{1/4}$--- which governs the evolution for $2 \lesssim t/\tauinv \lesssim 4$ (compare figures \ref{sub-fig:streamwise_vorticity_laminar_scaling_BCC} and \ref{sub-fig:streamwise_vorticity_turbulent_scaling_BCC}).
Conversely, the turbulent scaling does not apply for the HCP and the FCC squared streamwise vorticity (see figures \ref{sub-fig:streamwise_vorticity_turbulent_scaling_HCP} and \ref{sub-fig:streamwise_vorticity_turbulent_scaling_FCC}). 

The above turbulent scaling can be deduced in two different ways, namely: \textit{wall}- and \textit{Kolmogorov}-scaling approaches, by which two slightly different physical interpretations can be deduced.
The wall-scaling approach is based on approximations of the wall friction velocity $u_\tau$ and the corresponding friction length scale $\delta_\nu$, which are the crucial parameters to characterise the wall-bounded turbulent flows, such as plane Poiseuille and Hagen-Poiseuille flows.
This approach is inspired by the existence of the aforementioned unobstructed passages inside the BCC sphere pack (cf. \S \ref{sub-sec:dataset_BCC}), where the flow transitions into a turbulent-like state at late times when Hagen (and Reynolds) number is sufficiently high.
In fact, the considered parameter range is high enough to expect the emergence of the pore-scale turbulence after the aforementioned transient phase in general \citep{Wood2020}.  
Consequently, we approximate $u_\tau$ by the late-time velocity scaling in equation \eqref{eqn:superficial_velocity_scaling}, $u_{\tau} \sim \sqrt{d\,|\Vec{\nabla}\!\intrinsicavg{p}|/\rho}$, and using the definition of the length scale $\delta_{\nu} = \nu / u_{\tau}$, resulting:
\begin{equation}
     \omega_x\,\tauinv \sim \frac{d\,|\Vec{\nabla}\!\intrinsicavg{p}|}{\rho\,\nu}\,\tauinv \sim \frac{\tau_{\mathrm{visc}}}{\tauinv} \sim \Hg^{1/2} \ .
\end{equation}
The above estimation of velocity is consistent with the hydraulic-radius-based estimate given in \citet[][\S 2.3]{He2019}.
In contrast, the Kolmogorov-scaling is based on the potential flow velocity scale $U\sim |\Vec{\nabla}\!\intrinsicavg{p}| \tauinv/\rho$, and Kolmogorov length scale $\eta \sim (\nu^2/(2\mathsfbi{S}:\mathsfbi{S}))^{1/4}\sim (\nu^2/\omega_x^2)^{1/4}$, leading to:
\begin{equation}
      \omega_x\,\tauinv \sim \frac{\Vec{\nabla}\!\intrinsicavg{p}| \tauinv}{\rho} \left(\frac{\omega_x^2}{\nu^2}\right)^{1/4}\,\tauinv \sim \frac{1}{\tauinv} \left(\omega_x^2\,\tau_{\mathrm{visc}}^2\right)^{1/4}\,\tauinv \sim \left(\omega_x^2\,\tau_{\mathrm{visc}}^2\right)^{1/4}\ .
\end{equation}
Solving the above for $\omega_x$ results once again:
\begin{equation}
       \omega_x\,\tauinv\sim \frac{\tau_{\mathrm{visc}}}{\tauinv} \sim \Hg^{1/2}\ .
\end{equation}
Based on the above two approaches, somewhat different physical interpretations can be made: at late times in BCC, vorticity is governed by viscosity and the wall shear stress; or vorticity is governed by the small dissipative scales.

\subsection{Discussions}

By now, we have established that the emergence of nonlinearity is timed by the inviscid time scale $\tauinv$ in the limit of high Hagen numbers in all three sphere packs.
More specifically, we have shown, according to the streamwise vorticity on a cross-flow plane embodying a manifestation of nonlinearity, that a virtually identical evolution of flow structures is observed in the temporal range $0.6\le t/\tauinv \le 1.4$ in the high Hagen number limit.
This finding extends the results of \citet{Sakai2020} to the two additional sphere packs, and assigns a time scale to the consistent pore-scale flow structure development.
It is counter-intuitive in a sense that the vorticity development, which exclusively emerges on the sphere surfaces where the viscous influence should be predominant, is in fact controlled by an interplay with the neighbouring inviscid processes and their governing time scale.

In \S \ref{sub-sec:streamwise_vorticity_magnitude}, we have observed for small to intermediate inviscid times ($t/\tauinv \lesssim 2$), the vorticity magnitude approaches a laminar boundary layer scaling with $\omega \tauinv \Hg^{-1/4} \to \mathrm{const.}$
The observed scaling strongly indicates that the nonlinearity development at high Hagen numbers can be described by the boundary layer theory. 
This finding is complemented by our earlier result on the macroscopic velocity development (cf. \S \ref{sub-sec:nonlinear}) that the boundary layer theory, which is represented by the small-time asymptotics \eqref{eqn:small-time}, also describes the initial development of the superficial velocity $\superficialavg{u}$.
The growing nonlinearity (inertial influence) eventually forces the laminar boundary layer to be detached from the sphere surfaces (cf. \S \ref{sub-sec:streamwise_vorticity_dist}).

As mentioned repeatedly, the co-existence of irrotational outer flow and thin boundary (finite vorticity) layers growing via viscous diffusion is a general feature of the start-up flow in contact with no-slip boundaries. 
A canonical example which is relevant to the current flow cases at the pore-scale is the start-up flow around a circular cylinder, where the bluff body is started impulsively from rest, either to a fixed velocity or accelerated at a uniform rate. 
\cite{Honji1969} experimentally investigated both configurations, and their visualisations clearly show thin boundary layers on the cylinder surface growing under the potential core flow.
The growing boundary layers eventually separate from the cylinder surface around the rear stagnation point \citep{VanDommelen1982}, and form a pair of symmetrical vortices, which extends in the downstream with time.
In the limit of the considered parameters, the above vortical flow evolution was shown to be a universal feature in the both configurations.  
\cite{Honji1969} characterised the uniformly accelerating flow with dimensionless acceleration and time.
Note that their definition of the dimensionless acceleration ($=d^3a/\nu^2$, where $a$ is dimensional acceleration) is equivalent to the Hagen number in the current study, provided that our flow initially exhibits a constant acceleration ($a \sim |\Vec{\nabla}\!\intrinsicavg{p}|/\rho$), whereas their dimensionless time is equivalent to our $\tauvisc$.
The study showed that the length of the developing wake region is rather controlled by $a t^2/d$, which is essentially an inviscid dimensionless time.
A subsequent analytical and numerical investigation by \citet[][FIGURE 1]{collins_dennis_1974} revealed that the separation time under a similar acceleration-based inviscid normalisation becomes asymptotically constant for large Hagen numbers.
Since such flow separation is an inertial (nonlinear) process, their findings in the accelerating bluff body flow is consistent with our observations in the pore-scale flow evolution, especially in the context of the emergence of nonlinearity with respect to $\tauinv$.

For BCC, we have found the emergence of a turbulent scaling $\omega \tauinv \Hg^{-1/2} \to \mathrm{const.}$ for later times ($2 \lesssim t/\tauinv \lesssim 4$), whereas for the HCP and FCC geometries this scaling has no obvious significance.
A possible cause for this qualitative difference in the late-time scaling property between these two sphere pack groups is the unique geometric feature in the BCC, which has relatively large straight pore channels. 
In contrast, the main pore passages in the FCC and HCP sphere packs are obstructed by the sphere contact points, resulting in split-and-merge geometric patterns.
This conjuncture is consistent with the observation made for the H11 case in \citet[][Fig. 15]{Sakai2020} that the turbulent kinetic energy is highly concentrated in the wake regions behind the sphere contact points.
Those wake regions with high turbulent kinetic energy are enclosed by the high-momentum jets which are separated from the sphere surfaces at the contact points.
The existence of the high turbulent kinetic energy regions being not within, but adjacent to the high-momentum jets being created by the sphere contact points, is also visible in the steady-state FCC data of \cite[][FIGURE 3]{He2019}.
Therefore, it is safe to assume that the transition mechanism and distribution of turbulence in BCC differ significantly from the other two sphere packs, due to the absence of these contact points along the main pore passages, leading to the distinctive scaling property.  
Note that the sudden drops of the superficial velocity after extended periods of time (cf. \S \ref{sub-sec:nonlinear}) can also be attributed to this presumed difference in turbulence generation mechanism due to the geometric difference.
Despite the above difference during or after the emergence of turbulence, the generality of our main findings in this study, namely similarities in the nonlinearity developments across different sphere packs, should not be affected.
This is based on our understanding that the nonlinearity arises from the sphere surface boundary layers, where for $t\ll \tauinv$ the inertial convective term is much smaller than all other terms.

In summary, we confirmed the significance of the inviscid time scale stems from the fact that $\tauinv$ controls the formation time of the nonlinear pore-scale flow structures.
Moreover, it was shown both qualitatively and quantitatively that the pore-scale flow evolution becomes similar under the normalisation based on the laminar boundary-layer theory and $\tauinv$ in the high-$\Hg$ limit.
As for the transient linear flow, the initial development of the high-$\Hg$ flow at small times ($t \ll \tauinv$) is dictated by the laminar boundary layer growing under the potential core flow.
In the large time limit ($t \gg \tauinv$), on the other hand, the flow state is determined by the Reynolds number, which is a ratio between the viscous diffusion and the advective times, as in the steady-state.
In between these two limiting behaviours, when the flow is still accelerating, the above pore-scale similarity develops as an intermediate state.
During this intermediate period, the boundary layers thicken, subsequently detach from the sphere surface and migrate to and interact with the core flow.
The appearance of such similar sequence is not determined by a certain velocity magnitude or a time-scale ratio, but by a certain time.
Eventually, the flow state departs from the boundary-layer theory description and turbulence takes over the control.

From a practical viewpoint, the above findings imply that it is essential to incorporate the pore-scale boundary-layer behaviours in order to model the acceleration-dominated unsteady porous media flow accurately over the small and the intermediate times.
Such new modelling approach complements the existing model portfolio in which the linear and the weakly accelerating flow can be reasonably accurately approximated already \citep{Unglehrt2024}. 
A model for unsteady flows through porous media proposed in \citet[][pp.93f.]{lukas_phd_thesis} is a primary example, in which the linear flow model based on the dynamic permeability \citep{Pride.1993} for small Hagen numbers, and a steady laminar boundary layer scaling behaviour for high Hagen numbers were blended according to the method by \citet{Churchill1972}.
The performance of the new model was evaluated for oscillatory flows through HCP sphere pack driven by a sinusoidal time-dependent forcing, and it showed a significant improvement generally in the linear flow regime, as well as for low- and medium-frequency nonlinear flows \citep{lukas_phd_thesis}.

\section{Conclusion} 
\label{sec:conclusion}

We have investigated the start-up of flows through ordered porous media using high-fidelity direct numerical simulations in three canonical sphere packings: HCP, FCC, and BCC. By combining dimensional analysis, order-of-magnitude reasoning, and pore-scale diagnostics, we identified the characteristic time scales that govern the temporal development of these flows. In particular, we demonstrated that the inviscid time scale $\tauinv$ provides a universal measure for the onset of nonlinearity across all configurations and over a wide range of Hagen numbers. This contrasts with the conventional view that critical Reynolds numbers determine the emergence of nonlinear behaviour. Our results clearly show that in strongly accelerated flows, nonlinearity appears at a definite time, rather than at a definite velocity.

At the pore scale, the transient flow development is characterised by a thin Stokes boundary layer beneath an irrotational potential core. As time progresses, inertial effects within this boundary layer grow, reaching parity with pressure and viscous terms at $t \sim \tauinv$. This initiates the detachment of vorticity layers and the formation of inertial cores, marking the true onset of nonlinear dynamics. Despite differences in geometry and porosity, the sequence of events is remarkably similar among the three sphere packs, confirming the generality of the mechanism. The observed scaling of vorticity intensity with $\Hg^{1/4}$ further underpins the laminar boundary-layer origin of the process, whilst in the BCC case a transition towards turbulent-type scaling suggests the onset of pore-scale turbulence under sufficiently strong acceleration.

The present findings have direct implications for modelling strategies. First, they highlight the inadequacy of quasi-steady Reynolds-number-based parametrisations in capturing unsteady, acceleration-dominated regimes. Second, they demonstrate that inviscid scaling provides a simple yet robust framework to characterise the timing of nonlinear effects and velocity overshoots in start-up flows. Incorporating this knowledge into reduced-order or closure models will be crucial for reliable prediction of unsteady porous media dynamics in natural and engineering contexts, such as storm-driven aquifers, carbon storage operations, or subsurface contaminant transport.

Finally, the results point towards several open questions. 
Whilst the universality of $\tauinv$ has been established for ordered sphere packs, its extension to random or polydisperse media remains to be tested. 
Similarly, the precise mechanisms by which inertial cores and vortical structures interact to trigger turbulence in highly porous geometries deserve further exploration. Future studies, combining DNS with theoretical modelling and targeted experiments, will be essential to develop a comprehensive picture of transient porous media flows across the full range of operating conditions.

\vspace{1.5cm}

\noindent{\bf  Supplementary data\bf{.}} \label{SupMat} Time-resolved data of the superficial volume-averaged velocity were submitted with this manuscript. \\

\noindent{\bf Acknowledgements\bf{.}} The authors thank the anonymous reviewers of the earlier manuscript version (initially submitted to Phys. Rev. Fluids) for their valuable criticisms.
During preparation of this manuscript, the corresponding author used DeepL and ChatGPT (OpenAI) solely to identify potential English-language issues (e.g. grammar, phrasing, and readability) and to suggest alternative wording. 
These tools were not used to generate scientific content, analyse data, create figures, or draw conclusions. 
All changes were reviewed and approved by the corresponding author, who takes full responsibility for the manuscript content.\\

\noindent{\bf Funding\bf{.}} The authors gratefully acknowledge the financial support of the DFG under grant no. MA2062/13-1. Computing time was granted by the Leibniz Supercomputing Centre on its Linux-Cluster.\\

\noindent{\bf Declaration of Interests}. The authors report no conflict of interest. \\

\noindent{\bf  Author ORCID\bf{.}}  
Y. Sakai, \url{https://orcid.org/0000-0001-8125-6566}; \\ 
L. Unglehrt, \url{https://orcid.org/0000-0002-1299-0430}; \\
M. Manhart, \url{https://orcid.org/0000-0001-7809-6282}\\

\noindent{\bf Author contributions\bf{.}} 
All authors have contributed to the study concept and the definition of the parameter space.
The simulations and the data analysis were performed by YS. 
A first draft of the manuscript was written by YS.
The order-of-magnitude analysis was designed and performed by LU.
All authors read and approved the final manuscript.\\

\appendix
\section{Comparison of DNS datasets to previous studies}
\label{app:dnsdata}

Here, we compare the current DNS datasets to the results of the other analytical and DNS studies in literature, in order to assess the simulation quality. 

\subsection{Face-centred cubic packing (FCC)}
\label{app:dnsdata_fcc}

The dimensionless steady-state drag force $F$ is defined as:
\begin{equation}
    F = \frac{\left\vert\bnabla\!\intrinsicavg{p}\right\vert d^2}{18  \mu (1-\porosity) \overline{\superficialavg{u}}} = \frac{\Hg}{18  (1-\porosity) \Rey_{\mathrm{steady}}}\ .
\end{equation}

The dimensionless drag force from the linear flow simulation (case F1, $F=433$) agrees very
well with the analytical values from \cite{Zick1982} at $435$, and from \cite{Sangani1982a} at $438$.
The normalised Darcy permeability determined by the above linear flow case ($K/d^2=1.73 \times 10^{-4}$) also shows an excellent agreement with the value determined by \cite{Chapman1992} by means of a spectral method at $1.74 \times 10^{-4}$.

The steady-state properties of the nonlinear cases can be compared against the LBM simulation data of \cite{HILL2002a}, in which they provides an empirical linear fit of the $Re_\steady$-dependent dimensionless drag force for $20<\Rey_\steady<160$ \citep[cf.][equation 10]{HILL2002a}, for $\Rey_\steady>160$ \citep[cf.][equation 11]{HILL2002a}.
As shown in table \ref{tab:fcc_F_comparison}, the current dataset generally agrees well with the empirical predictions.

\begin{table}
    \centering
    \begin{tabular}{c c c}
        CASE & $F$ & $F_\mathrm{empirical}$ \\
        \hline
        F3 &588&566\\  
        F4 &661&633\\
        F5 &724&691\\
        F6 &897&892\\
        F7 &1188&1160\\
        F8 &1565&1553\\
        F9 &1867&1834\\
        F10&2086&2100\\
        F11&2292&2325\\
    \end{tabular}
    \caption{Comparison between the current FCC dataset and the empirical fit in \citet{HILL2002a}}
    \label{tab:fcc_F_comparison}
\end{table}

In \cite{HILL2002a}, it was reported that the streamwise superficial velocity of the FCC flow exhibits sinusoidal oscillations via Hopf bifurcation once the flow Reynolds number exceeds a critical value, which they estimated at $\Rey_\steady=57.4$.
If $\Rey_\steady$ is further increased above approximately $90$, the cross-flow components of the superficial
velocity also start to oscillate, which eventually leads to chaotic oscillations.
In the current dataset, both the primary and the secondary bifurcations occur in between F5 and F6, as the case F6 exhibits sinusoidal oscillations in all directions, whilst the superficial velocities in case F7 fluctuate chaotically (not shown).
The normalised amplitudes of the velocity fluctuations around the mean values are respectively: $\{A_{||},A_\perp\}=\{0.72,0.05\}$ for F6, which are in a great agreement with \citet[][FIGURE 2]{HILL2002a}, as well as an empirical fit provided in the same paper (cf. \citet[][equation (8)]{HILL2002a}, which predicts $A_\perp = 0.77$ for F6).
This finite level of $A_\perp$ indicates that the $\pi/2$ rotational symmetry on the cross-flow plane is broken in F6, whilst the same symmetry is preserved up to F5.

\subsection{Body-centred cubic packing (BCC)} \label{app:dnsdata_bcc}

As for the FCC dataset, the Stokes flow case (B1) was used to validate the simulation by determining the dimensionless drag force at $F=162.8$, 
which is in an excellent agreement with
the analytical values from \cite{Zick1982} at $163$, and from \cite{Sangani1982a} at $162$.
The corresponding normalised Darcy permeability from the linear flow data $K/d^2=5.16 \times 10^{-4}$ also compares well with the value of \cite{Chapman1992} at $5.02 \times 10^{-4}$.

We categorise B1 to be in the linear flow regime; the cases B2 through B5 are in the steady nonlinear flow regime since they all preserve $\pi/2$ rotational symmetry and are non-oscillatory.
The symmetry eventually breaks in B6 although the flow remains steady, which is a peculiar difference to the FCC cases.
For the sake of following the conventions in literature, however, we still assign B6 to the steady nonlinear flow.
In B7 we observe first signs of oscillations together with the aforementioned symmetry breaking (unsteady nonlinear), whilst B8--11 belong to the turbulent/chaotic flow category due to the development of chaotic oscillations even over the transient phase.

Note that the above two-stage symmetry breaking process being observed between B5 and B6 and between B6 and B7 is consistent with the one reported in \cite{Forslund2023}.
In their steady-state simulations with the BCC sphere pack, albeit at a higher $\porosity$ of $0.544$, the first transition point was determined at $\Rey_\mathrm{H}=179$, which corresponds to B6 in the current database, whilst their flow exhibits the second transition around $\Rey_\mathrm{H}=286$ corresponding to the gap between B6 and B7 in our simulations. 
This agreement, in spite of the larger porosity and therefore the absence of touching points in the simulations of \cite{Forslund2023}, indicates a robustness of the observed symmetry breaking process.

\section{Universality of small-time asymptotics}
\label{app:small-time}

Here, we show that the small-times asymptotics \eqref{eqn:small-time} are universal in the Darcy normalisation $\tauen$ and $\superficialavg{u}/(-\permeability \nabla\!\intrinsicavg{p}/\mu)$.

First, we divide equation \eqref{eqn:small-time} by the Darcy velocity and $\rho$:

\begin{equation}
    \frac{\mathrm{d}}{\mathrm{d}t}\left( \frac{\superficialavg{\Vec{u}} \mu}{-\permeability \nabla\!\intrinsicavg{p} } \right)  = 
    -\sqrt{\nu} \frac{2}{\Lambda} \int_{0}^{t} \frac{\mathrm{d}}{\mathrm{d}\tau} \left( \frac{\superficialavg{\Vec{u}} \mu}{-\permeability \nabla\!\intrinsicavg{p} } \right) \frac{1}{\sqrt{\pi(t-\tau)}}\,\mathrm{d}\tau
    +\frac{\nu}{\permeability}\frac{\porosity}{\alpha_\infty}
    \ .
    \label{eqn:small-time-step1}
\end{equation}

Then we multiply \eqref{eqn:small-time-step1} by $\tauen$, change the differentiation variable to $\hat{t} = t/\tauen$ and the integration variable to $\hat{\tau} = \tau/\tauen$:

\begin{equation}
    \frac{\mathrm{d}}{\mathrm{d}\hat{t}}\left( \frac{\superficialavg{\Vec{u}} \mu}{-\permeability \nabla\!\intrinsicavg{p} } \right) 
    = 
    - \frac{2}{\Lambda} \sqrt{\frac{\alpha_0 \permeability}{\porosity}}
    \int_{0}^{\hat{t}} \frac{\mathrm{d}}{\mathrm{d}\hat{\tau}}\left( \frac{\superficialavg{\Vec{u}} \mu}{-\permeability \nabla\!\intrinsicavg{p} } \right) \frac{1}{\sqrt{\pi(\hat{t}-\hat{\tau})}}\,\mathrm{d}\hat{\tau} 
    +\frac{\alpha_0}{\alpha_\infty}
    \ .
    \label{eqn:small-time-visc}
\end{equation}
Since the above equation does not feature any free dimensionless parameters, the corresponding solution is universal under this normalisation.

\bibliographystyle{jfm}
\bibliography{library}

@Article{	  whitaker:1996,
  author	= {Whitaker, S.},
  journal	= {Transp. Porous Media},
  number	= {1},
  pages		= {27--61},
  publisher	= {Springer},
  title		= {The {F}orchheimer equation: a theoretical development},
  volume	= {25},
  year		= {1996}
}

@Article{	  lowe2008a,
  author	= {Lowe, Ryan J. and Shavit, Uri and Falter, James L. and
		  Koseff, Jeffrey R. and Monismith, Stephen G.},
  doi		= {10.4319/lo.2008.53.6.2668},
  isbn		= {0024-3590},
  issn		= {00243590},
  journal	= {Limnol. Oceanogr.},
  month		= {nov},
  number	= {6},
  pages		= {2668--2680},
  title		= {{Modeling flow in coral communities with and without
		  waves: A synthesis of porous media and canopy flow
		  approaches}},
  url		= {http://doi.wiley.com/10.4319/lo.2008.53.6.2668},
  volume	= {53},
  year		= {2008}
}

@Article{	  rogers2016,
  author	= {Rogers, Justin S. and Monismith, Stephen G. and Koweek,
		  David A. and Torres, Walter I. and Dunbar, Robert B.},
  doi		= {10.1002/lno.10365},
  issn		= {19395590},
  journal	= {Limnol. Oceanogr.},
  number	= {6},
  pages		= {2191--2206},
  title		= {{Thermodynamics and hydrodynamics in an atoll reef system
		  and their influence on coral cover}},
  volume	= {61},
  year		= {2016}
}

@InCollection{	  dybbs1984,
  address	= {Dordrecht},
  author	= {Dybbs, A. and Edwards, R. V.},
  booktitle	= {Fundam. Transp. Phenom. Porous Media},
  doi		= {10.1007/978-94-009-6175-3{\_}4},
  isbn		= {978-94-009-6177-7},
  issn		= {1098-6596},
  month		= {dec},
  number	= {12},
  pages		= {199--256},
  pmid		= {25246403},
  publisher	= {Springer Netherlands},
  title		= {{A New Look at Porous Media Fluid Mechanics — Darcy to
		  Turbulent}},
  url		= {http://link.springer.com/10.1007/978-94-009-6175-3{\_}4
		  http://www.ncbi.nlm.nih.gov/pubmed/25246403
		  http://www.pubmedcentral.nih.gov/articlerender.fcgi?artid=PMC4249520},
  volume	= {58},
  year		= {1984}
}

@Article{	  hill2001,
  author	= {Hill, Reghan J. and Koch, Donald L. and Ladd, Anthony J.
		  C.},
  doi		= {10.1017/S0022112001005948},
  isbn		= {0022-1120},
  issn		= {0022-1120},
  journal	= {J. Fluid Mech.},
  pages		= {213--241},
  title		= {{The first effects of fluid inertia on flows in ordered
		  and random arrays of spheres}},
  url		= {http://www.journals.cambridge.org/abstract{\_}S0022112001005948},
  volume	= {448},
  year		= {2001}
}

@Article{	  hill2001a,
  author	= {Hill, Reghan J. and Koch, Donald L. and Ladd, Anthony J.
		  C.},
  doi		= {10.1017/S0022112001005936},
  isbn		= {0022-1120},
  issn		= {0022-1120},
  journal	= {J. Fluid Mech.},
  month		= {dec},
  pages		= {243--278},
  title		= {{Moderate-Reynolds-number flows in ordered and random
		  arrays of spheres}},
  url		= {http://www.journals.cambridge.org/abstract{\_}S0022112001005948
		  http://www.journals.cambridge.org/abstract{\_}S0022112001005936},
  volume	= {448},
  year		= {2001}
}

@Article{	  hill2002a,
  author	= {Hill, Reghan J. and Koch, Donald L.},
  doi		= {10.1017/S0022112002008947},
  issn		= {0022-1120},
  journal	= {J. Fluid Mech.},
  pages		= {59--97},
  title		= {{The transition from steady to weakly turbulent flow in a
		  close-packed ordered array of spheres}},
  url		= {http://www.journals.cambridge.org/abstract{\_}S0022112002008947},
  volume	= {465},
  year		= {2002}
}

@Article{	  chorin1968,
  author	= {Chorin, A. J.},
  doi		= {10.1023/A:1021973025166},
  issn		= {1572-9125},
  journal	= {Math. Comput.},
  number	= {3},
  pages		= {490--507},
  title		= {{Numerical Solution of the Navier-Stokes Equation}},
  url		= {http://www.ams.org/journals/mcom/1968-22-104/S0025-5718-1968-0242392-2/S0025-5718-1968-0242392-2.pdf},
  volume	= {42},
  year		= {1968}
}

@Article{	  williamson1980,
  author	= {Williamson, J. H.},
  doi		= {10.1016/0021-9991(80)90033-9},
  isbn		= {0021-9991},
  issn		= {10902716},
  journal	= {J. Comput. Phys.},
  number	= {1},
  pages		= {48--56},
  title		= {{Low-storage Runge-Kutta schemes}},
  volume	= {35},
  year		= {1980}
}

@Article{	  peller2006,
  title		= {High-Order Stable Interpolations for Immersed Boundary
		  Methods},
  author	= {Peller, Nikolaus and Duc, Anne Le and Tremblay,
		  Fr{\'e}d{\'e}ric and Manhart, Michael},
  year		= {2006},
  month		= dec,
  volume	= {52},
  pages		= {1175--1193},
  issn		= {02712091, 10970363},
  doi		= {10.1002/fld.1227},
  journal	= {Int. J. Numer. Methods Fluids},
  number	= {11}
}

@PhDThesis{	  peller2010,
  title		= {Numerische {{Simulation}} Turbulenter {{Str\"omungen}} Mit
		  {{Immersed Boundaries}}},
  author	= {Peller, Nikolaus},
  year		= {2010},
  month		= jan,
  address	= {{M\"unchen}},
  school	= {Technische Universit\"at M\"unchen}
}

@InProceedings{	  sakai2019,
  author	= {Sakai, Yoshiyuki and Mendez, Sandra and Strandenes,
		  H\r{a}kon and Ohlerich, Martin and Pasichnyk, Igor and
		  Allalen, Momme and Manhart, Michael},
  title		= {Performance Optimisation of the Parallel {CFD} Code
		  {MGLET} across Different {HPC} Platforms},
  year		= {2019},
  isbn		= {9781450367707},
  publisher	= {Association for Computing Machinery},
  address	= {New York, NY, USA},
  url		= {https://doi.org/10.1145/3324989.3325716},
  doi		= {10.1145/3324989.3325716},
  booktitle	= {Proceedings of the Platform for Advanced Scientific
		  Computing Conference},
  articleno	= {6},
  numpages	= {13},
  location	= {Zurich, Switzerland},
  series	= {PASC '19},
  number	= {}
}

@Article{	  sakai2020,
  author	= {Sakai, Y. and Manhart, M.},
  doi		= {10.1007/s10494-020-00168-4},
  journal	= {Flow, Turbul. Combust.},
  title		= {{Consistent flow structure evolution in accelerating flow
		  through hexagonal sphere pack}},
  year		= {2020}
}

@InProceedings{	  sakai2022,
  address	= {Osaka},
  author	= {Sakai, Yoshiyuki and Manhart, Michael},
  booktitle	= {12th Int. Symp. Turbul. Shear Flow Phenomena, TSFP 2022},
  number	= {1},
  pages		= {1--6},
  title		= {{Why Velocity Overshoots in Accelerating Porous Media
		  Flow}},
  url		= {http://www.tsfp-conference.org/proceedings/2022/255.pdf},
  year		= {2022}
}

@Article{	  he2019,
  author	= {He, Xiaoliang and Apte, Sourabh V. and Finn, Justin R. and
		  Wood, Brian D.},
  doi		= {10.1017/jfm.2019.403},
  issn		= {14697645},
  journal	= {J. Fluid Mech.},
  pages		= {608--645},
  title		= {{Characteristics of turbulence in a face-centred cubic
		  porous unit cell}},
  year		= {2019}
}

@Article{	  chapman1992,
  author	= {Chapman, A. M. and Higdon, J. J.L.},
  doi		= {10.1063/1.858507},
  issn		= {08998213},
  journal	= {Phys. Fluids A},
  number	= {10},
  pages		= {2099--2116},
  title		= {{Oscillatory Stokes flow in periodic porous media}},
  volume	= {4},
  year		= {1992}
}

@Article{	  stone1968,
  author	= {Stone, Herbert L},
  doi		= {10.1137/0705044},
  eprint	= {0705044},
  journal	= {SIAM J. Numer. Anal.},
  number	= {3},
  pages		= {530--558},
  primaryclass	= {10.1137},
  title		= {{Multidimensional Partial Differential Equations}},
  volume	= {5},
  year		= {1968}
}

@Article{	  forchheimer1901,
  author	= {Forchheimer, P.},
  journal	= {Z. Ver. Deutsch Ing.},
  pages		= {1782--1788},
  title		= {{Wasserbeweguin durch Boden}},
  volume	= {45},
  year		= {1901}
}

@Article{	  zhu2014,
  author	= {Zhu, Tao and Waluga, Christian and Wohlmuth, Barbara and
		  Manhart, Michael},
  doi		= {10.1007/s11242-014-0326-3},
  issn		= {01693913},
  journal	= {Transp. Porous Media},
  number	= {1},
  pages		= {161--179},
  title		= {{A Study of the Time Constant in Unsteady Porous Media
		  Flow Using Direct Numerical Simulation}},
  volume	= {104},
  year		= {2014}
}

@Article{	  zhu2016,
  title		= {Oscillatory {{Darcy Flow}} in {{Porous Media}}},
  author	= {Zhu, Tao and Manhart, Michael},
  year		= {2016},
  month		= jan,
  journal	= {Transp. Porous Media},
  volume	= {111},
  number	= {2},
  pages		= {521--539},
  issn		= {0169-3913, 1573-1634},
  doi		= {10.1007/s11242-015-0609-3}
}

@PhDThesis{	  zhu2016a,
  author	= {Zhu, Tao},
  title		= {Unsteady porous-media flows},
  year		= {2016},
  school	= {Technische Universität München},
  pages		= {103},
  language	= {en},
  note		= {}
}

@Book{		  schlichting.2017,
  title		= {Boundary-{{Layer Theory}}},
  author	= {Schlichting, Hermann and Gersten, Klaus},
  year		= {2017},
  edition	= {9th ed. 2017},
  publisher	= {{Springer Berlin Heidelberg : Imprint: Springer}},
  address	= {{Berlin, Heidelberg}},
  doi		= {10.1007/978-3-662-52919-5},
  isbn		= {978-3-662-52919-5},
  lccn		= {620.1064}
}

@Article{	  johnson.1987,
  title		= {Theory of Dynamic Permeability and Tortuosity in
		  Fluid-Saturated Porous Media},
  author	= {Johnson, David L. and Koplik, Joel and Dashen, Roger},
  year		= {1987},
  month		= mar,
  volume	= {176},
  pages		= {379},
  issn		= {0022-1120, 1469-7645},
  doi		= {10.1017/S0022112087000727},
  journal	= {J. Fluid Mech.}
}

@Article{	  collins_dennis_1974,
  title		= {Symmetrical flow past a uniformly accelerated circular
		  cylinder},
  volume	= {65},
  doi		= {10.1017/S0022112074001480},
  number	= {3},
  journal	= {J. Fluid Mech.},
  publisher	= {Cambridge University Press},
  author	= {Collins, W. M. and Dennis, S. C. R.},
  year		= {1974},
  pages		= {461–480}
}

@Article{	  pride.1993,
  title		= {Drag Forces of Porous-Medium Acoustics},
  author	= {Pride, Steven R. and Morgan, Frank Dale and Gangi, Anthony
		  F.},
  year		= {1993},
  month		= mar,
  journal	= {Phys. Rev. B},
  volume	= {47},
  number	= {9},
  pages		= {4964--4978},
  issn		= {0163-1829, 1095-3795},
  doi		= {10.1103/PhysRevB.47.4964},
  langid	= {english}
}

@Article{	  nguyen2018,
  author	= {Nguyen, Thien and Kappes, Ethan and King, Stephen and
		  Hassan, Yassin and Ugaz, Victor},
  doi		= {10.1007/s00348-018-2583-3},
  isbn		= {0123456789},
  issn		= {07234864},
  journal	= {Exp. Fluids},
  number	= {8},
  pages		= {0},
  publisher	= {Springer Berlin Heidelberg},
  title		= {{Time-resolved PIV measurements in a low-aspect ratio
		  facility of randomly packed spheres and flow analysis using
		  modal decomposition}},
  url		= {http://dx.doi.org/10.1007/s00348-018-2583-3},
  volume	= {59},
  year		= {2018}
}

@Article{	  maier1998,
  author	= {Maier, R. S. and Kroll, D. M. and Kutsovsky, Y. E. and
		  Davis, H. T. and Bernard, Robert S.},
  doi		= {10.1063/1.869550},
  issn		= {10706631},
  journal	= {Phys. Fluids},
  number	= {1},
  pages		= {60},
  title		= {{Simulation of flow through bead packs using the lattice
		  Boltzmann method}},
  url		= {http://link.aip.org/link/PHFLE6/v10/i1/p60/s1{\&}Agg=doi},
  volume	= {10},
  year		= {1998}
}

@Article{	  wood2020,
  author	= {Wood, Brian D. and He, Xiaoliang and Apte, Sourabh V.},
  doi		= {10.1146/annurev-fluid-010719-060317},
  issn		= {00664189},
  journal	= {Annu. Rev. Fluid Mech.},
  pages		= {171--203},
  title		= {{Modeling Turbulent Flows in Porous Media}},
  volume	= {52},
  year		= {2020}
}

@Article{	  sangani1982a,
  author	= {Sangani, A. S. and Acrivos, A.},
  doi		= {10.1016/0301-9322(82)90047-7},
  issn		= {03019322},
  journal	= {Int. J. Multiph. Flow},
  number	= {4},
  pages		= {343--360},
  title		= {{Slow flow through a periodic array of spheres}},
  volume	= {8},
  year		= {1982}
}

@Article{	  zick1982,
  author	= {Zick, A. A. and Homsy, G. M.},
  doi		= {10.1017/S0022112082000627},
  issn		= {14697645},
  journal	= {J. Fluid Mech.},
  pages		= {13--26},
  title		= {{Stokes flow through periodic arrays of spheres}},
  volume	= {115},
  year		= {1982}
}

@Article{	  forslund2023,
  author	= {Forslund, T. O.M. and Larsson, I. A.S. and
		  Hellstr{\"{o}}m, J. G.I. and Lundstr{\"{o}}m, T. S.},
  doi		= {10.1007/s11242-023-01966-w},
  isbn		= {0123456789},
  issn		= {15731634},
  journal	= {Transp. Porous Media},
  publisher	= {Springer Netherlands},
  title		= {{Steady-State Transitions in Ordered Porous Media}},
  url		= {https://doi.org/10.1007/s11242-023-01966-w},
  year		= {2023}
}

@Article{	  suekane2003,
  author	= {Suekane, Tetsuya and Yokouchi, Yasuo and Hirai,
		  Shuichiro},
  doi		= {10.1002/aic.690490103},
  issn		= {00011541},
  journal	= {AIChE J.},
  number	= {1},
  pages		= {10--17},
  title		= {{Inertial flow structures in a simple-packed bed of
		  spheres}},
  volume	= {49},
  year		= {2003}
}

@Article{	  unglehrt2022,
  author	= {Unglehrt, Lukas and Manhart, Michael},
  doi		= {10.1017/jfm.2022.496},
  issn		= {14697645},
  journal	= {J. Fluid Mech.},
  pages		= {1--32},
  title		= {{Onset of nonlinearity in oscillatory flow through a
		  hexagonal sphere pack}},
  volume	= {944},
  year		= {2022}
}

@Article{	  unglehrt2023,
  title		= {Decomposition of the Drag Force in Steady and Oscillatory
		  Flow through a Hexagonal Sphere Pack},
  author	= {Unglehrt, Lukas and Manhart, Michael},
  year		= {2023},
  month		= nov,
  journal	= {Journal of Fluid Mechanics},
  volume	= {974},
  pages		= {A32},
  publisher	= {{Cambridge University Press}},
  issn		= {0022-1120, 1469-7645},
  doi		= {10.1017/jfm.2023.798},
  urldate	= {2023-11-09},
  langid	= {english}
}

@Article{	  forslund2021,
  title		= {Non-Stokesian flow through ordered thin porous media
		  imaged by tomographic-PIV},
  author	= {Forslund, Tobias OM and Larsson, IA Sofia and Lycksam,
		  Henrik and Hellstr{\"o}m, J Gunnar I and Lundstr{\"o}m, T
		  Staffan},
  journal	= {Experiments in Fluids},
  volume	= {62},
  pages		= {1--12},
  year		= {2021},
  publisher	= {Springer}
}

@Article{	  honji1969,
  author	= {Honji, Hiroyuki and Taneda, Sadatoshi},
  journal	= {Reports Res. Inst. Appl. Mech. Kyushu Univ.},
  number	= {59},
  pages		= {187--193},
  publisher	= {Research Institute for Applied Mechanics, Kyushu
		  University},
  title		= {{Time-dependent flow around a circular cylinder
		  accelerated uniformly from one steady speed to another}},
  volume	= {17},
  year		= {1969}
}

@InProceedings{	  vandommelen1982,
  author	= "Van Dommelen, L. L. and Shen, S. F.",
  editor	= "Cebeci, Tuncer",
  title		= "The Genesis of Separation",
  booktitle	= "Numerical and Physical Aspects of Aerodynamic Flows",
  year		= "1982",
  publisher	= "Springer Berlin Heidelberg",
  address	= "Berlin, Heidelberg",
  pages		= "293--311",
  isbn		= "978-3-662-12610-3"
}

@Article{	  horton2009,
  author	= {Horton, N. A. and Pokrajac, D.},
  title		= "{Onset of turbulence in a regular porous medium: An
		  experimental study}",
  journal	= {Physics of Fluids},
  volume	= {21},
  number	= {4},
  pages		= {045104},
  year		= {2009},
  month		= {04},
  issn		= {1070-6631},
  doi		= {10.1063/1.3091944},
  url		= {https://doi.org/10.1063/1.3091944}
}

@Article{	  unglehrt2024,
  author	= {Unglehrt, Lukas and Manhart, Michael},
  doi		= {10.1007/s11242-024-02110-y},
  issn		= {0169-3913},
  journal	= {Transp. Porous Media},
  month		= {jul},
  volume	= {151},
  issue		= {10-11},
  pages		= {2183-2213},
  publisher	= {Springer Netherlands},
  title		= {{Assessment of Models for Nonlinear Oscillatory Flow
		  Through a Hexagonal Sphere Pack}},
  year		= {2024}
}

@Article{	  santos2012,
  author	= {Santos, Isaac R. and Eyre, Bradley D. and Huettel,
		  Markus},
  doi		= {10.1016/j.ecss.2011.10.024},
  issn		= {02727714},
  journal	= {Estuar. Coast. Shelf Sci.},
  month		= {feb},
  pages		= {1--15},
  publisher	= {Elsevier Ltd},
  title		= {{The driving forces of porewater and groundwater flow in
		  permeable coastal sediments: A review}},
  url		= {http://dx.doi.org/10.1016/j.ecss.2011.10.024
		  https://linkinghub.elsevier.com/retrieve/pii/S027277141100446X},
  volume	= {98},
  year		= {2012}
}

@Article{	  robinson2007,
  author	= {Robinson, C. and Li, L. and Barry, D.A.},
  doi		= {10.1016/j.advwatres.2006.07.006},
  issn		= {03091708},
  journal	= {Adv. Water Resour.},
  month		= {apr},
  number	= {4},
  pages		= {851--865},
  title		= {{Effect of tidal forcing on a subterranean estuary}},
  url		= {https://linkinghub.elsevier.com/retrieve/pii/S0309170806001217},
  volume	= {30},
  year		= {2007}
}

@Article{	  nordio2023,
  author	= {Nordio, Giovanna and Frederiks, Ryan and Hingst, Mary and
		  Carr, Joel and Kirwan, Matt and Gedan, Keryn and Michael,
		  Holly and Fagherazzi, Sergio},
  doi		= {10.1029/2022GL100191},
  issn		= {0094-8276},
  journal	= {Geophys. Res. Lett.},
  month		= {jan},
  number	= {1},
  pages		= {1--10},
  title		= {{Frequent Storm Surges Affect the Groundwater of Coastal
		  Ecosystems}},
  url		= {https://agupubs.onlinelibrary.wiley.com/doi/10.1029/2022GL100191},
  volume	= {50},
  year		= {2023}
}

@Article{	  sacconi2020,
  author	= {Sacconi, Andrea and Mahgerefteh, Haroun},
  doi		= {10.1016/j.energy.2019.116530},
  issn		= {03605442},
  journal	= {Energy},
  pages		= {116530},
  publisher	= {Elsevier Ltd},
  title		= {{Modelling start-up injection of CO2 into highly-depleted
		  gas fields}},
  url		= {https://doi.org/10.1016/j.energy.2019.116530},
  volume	= {191},
  year		= {2020}
}

@Article{	  class2009,
  author	= {Class, Holger and Ebigbo, Anozie and Helmig, Rainer and
		  Dahle, Helge K. and Nordbotten, Jan M. and Celia, Michael
		  A. and Audigane, Pascal and Darcis, Melanie and Ennis-King,
		  Jonathan and Fan, Yaqing and Flemisch, Bernd and Gasda,
		  Sarah E. and Jin, Min and Krug, Stefanie and Labregere,
		  Diane and {Naderi Beni}, Ali and Pawar, Rajesh J. and Sbai,
		  Adil and Thomas, Sunil G. and Trenty, Laurent and Wei,
		  Lingli},
  doi		= {10.1007/s10596-009-9146-x},
  issn		= {1420-0597},
  journal	= {Comput. Geosci.},
  month		= {dec},
  number	= {4},
  pages		= {409--434},
  title		= {{A benchmark study on problems related to CO2 storage in
		  geologic formations}},
  url		= {http://link.springer.com/10.1007/s10596-009-9146-x},
  volume	= {13},
  year		= {2009}
}

@Article{	  graham_2019,
  title		= {Decomposition of the forces on a body moving in an
		  incompressible fluid},
  volume	= {881},
  doi		= {10.1017/jfm.2019.788},
  journal	= {Journal of Fluid Mechanics},
  author	= {Graham, W. R.},
  year		= {2019},
  pages		= {1097–1122}
}

@Book{		  conway_sphere_1999,
  address	= {New York, NY},
  series	= {Grundlehren der mathematischen {Wissenschaften}},
  title		= {Sphere {Packings}, {Lattices} and {Groups}},
  volume	= {290},
  copyright	= {http://www.springer.com/tdm},
  isbn		= {978-1-4419-3134-4 978-1-4757-6568-7},
  url		= {http://link.springer.com/10.1007/978-1-4757-6568-7},
  urldate	= {2025-10-30},
  publisher	= {Springer},
  author	= {Conway, J. H. and Sloane, N. J. A.},
  editor	= {Chern, S. S. and Eckmann, B. and De La Harpe, P. and
		  Hironaka, H. and Hirzebruch, F. and Hitchin, N. and
		  Hörmander, L. and Knus, M.-A. and Kupiainen, A. and
		  Lannes, J. and Lebeau, G. and Ratner, M. and Serre, D. and
		  Sinai, Ya. G. and Sloane, N. J. A. and Tits, J. and
		  Waldschmidt, M. and Watanabe, S. and Berger, M. and Coates,
		  J. and Varadhan, S. R. S.},
  year		= {1999},
  doi		= {10.1007/978-1-4757-6568-7}
}

@Article{	  smeulders_dynamic_1992,
  title		= {Dynamic permeability: reformulation of theory and new
		  experimental and numerical data},
  volume	= {245},
  copyright	= {https://www.cambridge.org/core/terms},
  issn		= {0022-1120, 1469-7645},
  shorttitle	= {Dynamic permeability},
  url		= {https://www.cambridge.org/core/product/identifier/S0022112092000429/type/journal_article},
  doi		= {10.1017/S0022112092000429},
  language	= {en},
  urldate	= {2025-11-27},
  journal	= {J. Fluid Mech.},
  author	= {Smeulders, D. M. J. and Eggels, R. L. G. M. and Van
		  Dongen, M. E. H.},
  month		= dec,
  year		= {1992},
  pages		= {211--227}
}

@PhDThesis{	  lukas_phd_thesis,
  author	= {Unglehrt, Lukas Maximilian},
  title		= {Oscillatory flow through porous media},
  year		= {2024},
  school	= {Technische Universität München},
  pages		= {337},
  language	= {en},
  note		= {},
  url		= {https://mediatum.ub.tum.de/1729913},
  doi		= {}
}

@Article{	  churchill1972,
  author	= {Churchill, S. W. and Usagi, R.},
  title		= {A general expression for the correlation of rates of
		  transfer and other phenomena},
  journal	= {AIChE Journal},
  volume	= {18},
  number	= {6},
  pages		= {1121-1128},
  doi		= {https://doi.org/10.1002/aic.690180606},
  url		= {https://aiche.onlinelibrary.wiley.com/doi/abs/10.1002/aic.690180606},
  eprint	= {https://aiche.onlinelibrary.wiley.com/doi/pdf/10.1002/aic.690180606},
  year		= {1972}
}
\end{document}